\documentclass[nofootinbib,prd,eqsecnum,showpacs,showkeys,preprintnumbers,altaffilletter]{revtex4} 

\usepackage{color}
\usepackage{float}
\usepackage{amsmath}
\usepackage{amssymb}
\usepackage{amsfonts}
\usepackage{graphicx}
\usepackage{enumerate}
\usepackage{textcomp}
\usepackage{esint}
\usepackage[normalem]{ulem} 
\usepackage{epstopdf}
\usepackage{tabularx}
\usepackage{multirow}
\usepackage[unicode=true,pdfusetitle,
 bookmarks=true,bookmarksnumbered=false,bookmarksopen=false,
 breaklinks=false,pdfborder={0 0 1},backref=false,colorlinks=false]
 {hyperref}
 
\makeatletter

\makeatletter

\renewcommand{\[}{\begin{equation}}
\renewcommand{\]}{\end{equation}}
\newcommand{\fnl}{f_{\rm NL}}

\makeatother

\begin{document}

\title{Non-Gaussianity in multiple three-form field inflation}

\author{K. Sravan Kumar $^{1,2}$}
\email{sravan@ubi.pt}

\author{{David J. Mulryne $^{3}$}}
\email{d.mulryne@qmul.ac.uk}

\author{{Nelson J. Nunes $^{4}$}}
\email{njnunes@fc.ul.pt}

\author{Jo\~ao Marto $^{1,2}$}
\email{jmarto@ubi.pt}

\author{Paulo Vargas Moniz $^{1,2}$}
\email{pmoniz@ubi.pt}

\date{\today}

\affiliation{ 
 $^{1}$Departamento de F\'{i}sica, Universidade da Beira Interior, 6200 Covilh\~a, Portugal\\
 $^{2}$Centro de Matem\'atica e Aplica\c{c}\~oes  da Universidade da Beira Interior (CMA-UBI), 6200 Covilh\~a, Portugal\\
 $^{3}${School of Physics and Astronomy, Queen Mary University of London, Mile End Road, London E1 4NS, United Kingdom}\\
 $^{4}${Faculdade de Ci\^encias, Instituto de Astrof\'{i}sica e Ci\^encias do Espa\c{c}o, Universidade de Lisboa, Campo Grande, PT1749-016 Lisboa, Portugal}
 }

\begin{abstract}
In this work, we present a method for implementing the $\delta N$
formalism to study the primordial non-Gaussianity produced in multiple
three-form field inflation. Using a dual description relating three-form
fields to noncanonical scalar fields, and employing existing results,
we produce expressions for the bispectrum of the curvature perturbation
in terms of three-form quantities. We study the bispectrum generated
in a two three-form field inflationary scenario for a particular potential
that for suitable values of the parameters was found in earlier work
to give values of the spectral index and ratio of tensor to scalar
perturbations compatible with current bounds. We calculate the reduced
bispectrum for this model, finding an amplitude in equilateral and
orthogonal configurations of ${\cal O}(1)$ and in the squeezed limit
of ${\cal O}(10^{-3})$. We confirm, therefore, that this three-form
inflationary scenario is compatible with present observational constraints. 
\end{abstract}

\keywords{Primordial non-Gaussianity, $\delta N$ formalism, multiple three-forms}

\pacs{98.80.Cq, 98.70.Vc}

\date{\today} 

\maketitle

\section{Introduction}

Inflation is a successful paradigm that solves the horizon and flatness
problems \cite{Guth:1980zm}. Measurements of the cosmic microwave
background (CMB) anisotropies confirm that the primordial density
perturbations are close to scale invariant, adiabatic and Gaussian
\cite{Ade:2015xua,Ade:2015lrj,Ade:2015ava}. This is expected from
the simplest models of inflation and confirms inflation as our favored
theory for the origin of structure. The results from Plank 2015 \cite{Ade:2015xua}
and the BICEP2/Keck Array and Planck joint analysis \cite{Ade:2015tva}
also severely constrain the amplitude of gravitational waves produced
by inflation, with the latest bounds on the tilt of the scalar power
spectrum $\left(n_{s}\right)$ and the tensor to scalar ratio $\left(r\right)$
given by \cite{Ade:2015lrj} 
\begin{equation}
n_{s}=0.968\pm0.006\quad\mathrm{and}\quad r_{0.002}<0.09\quad\text{at}\;95\%\;\text{C.L.}
\end{equation}

Despite its observational successes, however, there are a considerable
variety of different models of inflation that can be motivated theoretically
\cite{Martin:2014vha,Martin:2015dha}. These include multifield models,
and models with noncanonical scalar fields. The degeneracy of the
predictions from various models of inflation is an ongoing problem
for cosmologists. One way to probe further the nature of inflation
is to study the statistics of the perturbations it produces beyond
the two-point correlation function \cite{Tsujikawa:2014rta}, starting
with the three-point function. This is parametrized in Fourier space
by the bipsectrum \cite{Maldacena:2002vr,Seery:2005gb,Seery:2005wm,Chen:2006nt,Wands:2010af,Chen:2010xka},
a function of the amplitude of three wave vectors that sum to zero
as a consequence of momentum conservation. Although most canonical
single field models of inflation produce an unobservably small bispectrum,
multifield and noncanonical models in particular can produce levels
in tension with present or detectable by future probes. The former
produces a bispectrum of close to the \char`\"{}local shape\char`\"{}.
This is a function of three wave numbers that peaks in the squeezed
limit where two wave numbers are much larger than the third. The latter
produces a bispectrum of \char`\"{}equilateral shape\char`\"{} that
tends to zero in the squeezed limit, but peaks when all three wave numbers
are similar in size (see e.g. \cite{Chen:2010xka,Byrnes:2014pja}
for reviews). A third shape is often considered that peaks on folded
triangles, where two wave numbers are approximately half of the third,
and can be produced by models with non-Bunch-Davis initial conditions
\cite{Chen:2010xka}. Planck 2015 has put constraints on these shapes.
Introducing three parameters $\fnl^{{\rm loc}}$, $\fnl^{{\rm equi}}$
and $\fnl^{{\rm ortho}}$, which parametrize the overall amplitude
of a local, equilateral and orthonormal shape template for the bispectrum,
Planck 2015 tells us that 
\begin{equation}
f_{\rm NL}^{\rm local}=0.8\pm 5.0 \,,\quad f_{\rm NL}^{\rm equi}=-4\pm 43 \,,
\quad f_{\rm NL}^{\rm ortho}=-26\pm 21,%
\end{equation}
at $68\%$ C.L. (see Ref.~\cite{Ade:2013ydc}). These bounds are stringent,
though it is too early to exclude noncanonical or multifield inflationary
models by means of Gaussianity (see e.g. \cite{Elliston:2013afa,Baumann:2014cja,Vennin:2015vfa}.)

Despite the success of inflation driven by scalar fields, three-forms
provide a viable alternative (and a viable model of dark energy) \cite{Nunes-1,Nunes-2,GK09,Felice-1,Mulryne:2012ax,Koivisto:2012xm,Kumar:2014oka,Barros:2015evi}. {Inflation considering multiple three-form fields
has been investigated in the past as these models are
important due to their connection to string theory
scenarios \cite{Groh:2012tf}. Typically they have been written down with quadratic potentials but it useful to consider generalizations.}
In this article we therefore study how to calculate the bispectrum
in any multiple three-form inflationary scenario. To do so we develop
a method to adapt the $\delta N$ formalism \cite{Lyth:1984gv,Sasaki1995,Wands2000,Lyth2004,Lyth:2005du}
to the three-form setting. We then calculate the bispectrum generated
in a concrete model with two three-forms. Inflationary scenarios with
two three-forms were proposed in \cite{Kumar:2014oka}, and shown
under a suitable choice of the three-form potential and initial conditions
to satisfy the Planck data concerning the power spectrum and tensor
to scalar ratio. Here we compute the bispectrum for the successful
example considered in that paper and check that it is also consistent with
the latest observational constraints.

The plan of the paper is as follows. In Sec.\ref{N3formmodel}
we briefly summarize the $\mathbb{N}$ three-form inflationary model
studied in Ref. \cite{Kumar:2014oka}. Subsequently, in Sec. \ref{DeltaNformalism}
we discuss the bispectrum and describe a procedure to adapt the $\delta N$
formalism \cite{Lyth:2005fi} to multiple three-forms to calculate
it. We explain a numerical method for calculating derivatives of the
unperturbed number of $e$-foldings with respect to the unperturbed
three-form field values at sound horizon crossing, and show how these
derivatives can be related to those of a dual scalar field description.
In turn these can be used in combination with existing results to
compute the bispectrum. We stress that although our method utilizes
the dual scalar field description, it is not possible in general to
simply pass to that description and work solely with a scalar field
model. In Sec. \ref{NG23forms} we consider an explicit example
from Ref.~\cite{Kumar:2014oka} that provides a power-spectrum compatible
with Planck constraints and compute the bispectrum in that model.
We quantify and compare the momentum dependent contribution and momentum
independent contributions of the reduced bispectrum and plot the shape
of the bispectrum. We conclude in Sec.\ref{Concl}.

\section{Multiple three-form inflation}

\label{N3formmodel} In this section, we briefly present the {inflationary}
model \textit{\emph{with}} $\mathbb{N}$ three-form fields introduced
in \cite{Kumar:2014oka}. We take a flat Friedmann-Lema\^itre-Robertson-Walker
(FLRW) cosmology, described with the metric $ds^{2}=-dt{}^{2}+a^{2}(t)d\boldsymbol{x}^{2},$
where $a(t)$ is the scale factor with $t$ cosmic time. The general
action for two three-form fields minimally coupled to Einstein gravity
can be written as\footnote{We work in the units of Planck mass $M_{{\rm Pl}}=1$.}
\begin{equation}
S=-\int d^{4}x\sqrt{-g}\left[\frac{1}{2}R-\sum_{I=1}^{\mathbb{N}}\left(\frac{1}{48}F_{I}^{2}+V(A_{I}^{2})\right)\right]\,,\label{N3-form action}
\end{equation}
where $A_{\beta\gamma\delta}^{(I)}$ is the $I{\textrm{th}}$ three-form
field and squared quantities indicate contraction of all the indices.
The strength tensor of the three-form is given by\footnote{Throughout this article, the latin index ``I" will be used to refer to the
number of the quantity (e.g., the three-form field) or the $I{\textrm{th}}$
quantity/field. The other latin indices, which take the values $i,j=1,2,3...$,
will indicate the three-dimensional quantities; whereas the greek
indices are used to denote four-dimensional quantities and they
stand for $\mu,\nu=0,1,2,3$.} 
\begin{equation}
F_{\alpha\beta\gamma\delta}^{(I)}\equiv4\nabla_{[\alpha}A_{\beta\gamma\delta]}^{(I)},\label{N3f-Maxw-1}
\end{equation}
where antisymmetrisation is denoted by square brackets. As we have
assumed a homogeneous and isotropic universe, the three-form fields
depend only on time and hence only the spacelike components are dynamical. Therefore the nonzero components are given by \cite{Nunes-2}
\begin{equation}
A_{ijk}^{(I)}=a^{3}(t)\epsilon_{ijk}\chi_{I}(t)\quad\Rightarrow A_{I}^{2}=6\chi_{I}^{2},\label{NNZ-comp-1-1}
\end{equation}
where $\chi_{I}(t)$ is a comoving field associated to the $n$th
three-form field and $\epsilon_{ijk}$ is the standard three-dimensional
Levi-Civita symbol.

In general, any $p$-form in $d$ dimensions has a dual of $(d-p)-$form
\cite{GK09,Mulryne:2012ax}. In our case three-form field $\left(A\right)$
and its field tensor four-form $(F)$ are dual to a vector and a scalar
field respectively which can be expressed as \cite{Mulryne:2012ax}
\begin{equation}
A_{\mu\nu\rho}=\epsilon_{\alpha\mu\nu\rho}B^{\alpha}\,,\quad F_{\mu\nu\rho\sigma}=-\epsilon_{\mu\nu\rho\sigma}\phi\,,
\label{dual definitions}
\end{equation}
where $\epsilon_{\mu\nu\rho\sigma}$is an antisymmetric tensor.

The corresponding action for the scalar field dual representation
of the $\mathbb{N}$ three-forms is \cite{Kumar:2014oka,Mulryne:2012ax}
\begin{equation}
S=-\int d^{4}x\sqrt{-g}\left[\frac{1}{2}R+P\left(X,\phi_{I}\right)\right]\,,
\end{equation}
where 
\begin{equation}
P\left(X,\phi_{I}\right)=\sum_{I=1}^{\mathbb{N}}\left(\chi_{I}V_{I,\chi_{I}}-V\left(\chi_{I}\right)-\frac{\phi_{I}^{2}}{2}\right)\,,\label{dual action}
\end{equation}
with $X=-\frac{1}{2}G^{IJ}\left(\phi\right)\partial_{\mu}\phi_{I}\partial^{\mu}\phi_{J}$.
In this model the field metric is $G^{IJ}\left(\phi\right)=\delta^{IJ}$,
therefore we have $X=\sum X_{I}$. The three-from fields still present
on the right-hand side of Eq.~(\ref{dual action}) should be viewed
as functions of the kinetic terms $X_{I}$ though the inverse of the
relation 
\begin{equation}
X_{I}=\frac{1}{2}V_{,\chi_{I}}^{2}\,.\label{relation}
\end{equation}

Considering the {large amount of noncanonical scalar fields studies} in cosmology,
it might be tempting to think that given a three-form theory the best
way to proceed would be to simply pass to the dual scalar field theory
and work solely with scalar field quantities. {However, starting} from a set
of massive three-form fields {makes the task of analytically
writing the dual scalar field theory very difficult,} except for very particular potentials \cite{Mulryne:2012ax}.
{This can be seen by noting the technical difficulty found when one tries to invert Eq.~(\ref{relation}).}
Yet, in a similar manner to that advocated in Ref.~\cite{Mulryne:2012ax}
for the single field case, we see that we can still make use
of the dual theory indirectly.

For a background unperturbed FLRW cosmology, we can use the dualities
defined in Eq. (\ref{dual definitions}) to write the following relation
between a three-form field and its dual scalar field 
\begin{equation}
\phi_{I}=\dot{\chi}_{I}+3H\chi_{I}.\label{3-formdual}
\end{equation}
Moreover from action (\ref{N3-form action}) the background Klein-Gordon
equations for the $\mathbb{N}$ three-form fields read 
\begin{equation}
\ddot{\chi}_{I}+3H\dot{\chi}_{I}+3\dot{H}\chi_{I}+V_{,\chi_{I}}=0.\label{NDiff-syst-1-1}
\end{equation}
The Friedmann equations are 
\begin{equation}
H^{2}=\frac{1}{6}\left[\sum_{I=1}^{\mathbb{N}}\left(\dot{\chi}_{I}+3H\chi_{I}\right)^{2}+2V\right]\,,\label{NFriedm-1-1}
\end{equation}
and 
\begin{equation}
\dot{H}=-\frac{1}{2}\left[\sum_{I=1}^{\mathbb{N}}V_{,\chi_{I}}\chi_{I}\right]\,.\label{NHdot-1-1}
\end{equation}
We express the field equations of motion (\ref{NDiff-syst-1-1}) in
terms of $e$-folding time, $N=\ln\,a(t)$, as 
\begin{equation}
H^{2}\chi_{I}^{\prime\prime}+\left(3H^{2}+\dot{H}\right)\chi_{I}^{\prime}+3\dot{H}\chi_{I}+V_{,\chi_{I}}=0\,,\label{NDiff-syst-1-1-3}
\end{equation}
where $\chi{}_{I}^{\prime}\equiv d\chi_{I}/dN$. And the Hubble parameter
from Eq. (\ref{NFriedm-1-1}) can be expressed as 
\begin{equation}
H^{2}=\frac{V\left(\chi_{I}\right)}{3\left(1-\sum_{I}w_{I}^{2}\right)}\,.\label{hubble}
\end{equation}
where $w_{I}=\frac{\chi_{I}^{\prime}+3\chi_{I}}{\sqrt{6}}\,.$

{The three-form field equations (\ref{NDiff-syst-1-1-3})
can also be written in the autonomous form as \cite{Kumar:2014oka}
\begin{equation}
\chi_{I}^{\prime}=3\left(\sqrt{\frac{2}{3}}w_{I}-\chi_{I}\right),\label{Nauton-n-n}
\end{equation}
\begin{equation}
w_{I}^{\prime}  =  \frac{3}{2}\left(1-\sum_{I}w_{I}^{2}\right)
\left[\lambda_{I}\left(\chi_{I}w_{I}-\sqrt{\frac{2}{3}}\right)+
\overset{\mathbb{N}}{\underset{\underset{I\neq J}{J=1}}{\sum}\chi_{J}\lambda_{J}}\right],\label{Nwprime-n}
\end{equation}
where $\lambda_{n}=V_{,\chi_{I}}/V$.
The fixed points of the dynamical system (\ref{Nauton-n-n})-(\ref{Nwprime-n})
are 
\begin{equation}
\begin{aligned}\chi_{Ic}= & \sqrt{\frac{2}{3}}w_{I},\hspace{1cm}w_{Ic}=\frac{\lambda_{I}}{\sqrt{\sum_{I}\lambda_{I}^{2}}},\end{aligned}
\label{fixed-P2}
\end{equation}
Based on the analytical and numerical studies of two three-form
inflation, which is detailed in Ref. \cite{Kumar:2014oka}, we can have the following two
types of slow-roll inflationary scenarios (corresponding to the different
trajectories in $\mathbb{N}$ three-form fields space), 
\begin{itemize}
\item  Type I inflation: It precisely produces
straight line trajectories in field space, where all the three-form fields
driving inflation satisfy $\chi^{\prime}{}_{I}\approx0$. This scenario
shares some similarities with multiple scalar fields assisted inflation \cite{Liddle-1}. 
In this case, the three-form fields sit near their
respective fixed points (\ref{fixed-P2}) until the end of inflation. Subsequently,
they oscillate collectively at the potential minimum. 
\item  Type II inflation: It produces curved trajectories
in field space where all the three-form fields driving inflation satisfy
$\chi_{I}^{\prime}\not\approx0$. In this case, the three-form fields have the freedom
to slowly evolve away from their respective fixed points (\ref{fixed-P2})
until the end of inflation. Finally, and also in this scenario,
they oscillate collectively at the potential minimum. 
\end{itemize}
In Ref. \cite{Kumar:2014oka} it was explicitly shown
that the type I inflation does not produce any isocurvature perturbations
that may source the curvature perturbations on superhorizon scales.
Therefore, we naively expect negligible non-Gaussianities in the type
I scenario. Whereas in type II inflation, where the three-form fields present a
different dynamics, we can expect a significant signal of non-Gaussianities.
Therefore, in the present work we exclusively focus our attention on the type II
inflationary scenarios. }

In subsequent sections, our strategy { (based on the three-form duality) to calculate
non-Gaussianities will be to use equations derived for multiple scalar
fields. However, we express the quantities involved in terms of the three-form fields. }
In particular, we need the following derivatives, which we compute here for later use, 
\begin{equation}
P_{,X}\equiv\sum_{I}P_{,X_{I}}=\sum_{I}P_{,\chi_{I}}\left(\frac{\partial\chi_{I}}{\partial X_{I}}\right)=\sum_{I}\frac{\chi_{I}}{V_{\chi_{I}}}\,.\label{PX}
\end{equation}
And similarly 
\begin{eqnarray}
P_{,X_{I}X_{I}} & = & \frac{1}{V_{,\chi_{I}\chi_{I}}V_{,\chi_{I}}^{2}}-\frac{\chi_{I}}{V_{,\chi_{I}}^{3}}\,.\label{PX23}\\
P_{,X_{I}X_{I}X_{I}} & = & -\frac{V_{,\chi_{I}\chi_{I}\chi_{I}}}{V_{,\chi_{I}\chi_{I}}^{3}V_{,\chi_{I}}^{2}}+\frac{3\chi_{I}}{V_{,\chi_{I}}^{5}}-\frac{3}{V_{,\chi_{I}}^{4}V_{,\chi_{I}\chi_{I}}}\,.\label{PX33}\\
P_{,I} & = & -\phi_{I}=-\sqrt{6}Hw_{I}\,.\label{PI}
\end{eqnarray}

\section{Non-Gaussianity and the $\delta N$ formalism }

\label{DeltaNformalism}

\subsection{The $\delta N$ formalism }

\label{delta N formalism}



The $\delta N$ formalism is based on the separate universe assumption
\cite{Lyth:1984gv,Starobinsky:1986fxa,Sasaki1995,Wands2000,Lyth2004,Lyth:2005du}
and provides a powerful tool to evaluate the superhorizon evolution
of the curvature perturbation. In the case of multiple three-forms,
however, the direct implementation of the $\delta N$ formalism would
be cumbersome. Using the formal relation between three-forms and their
scalar field duals \cite{Mulryne:2012ax,Kumar:2014oka}, however,
one can indirectly implement the $\delta N$ formalism while still
employing only three-form quantities that are easy to calculate.

The $\delta N$ formalism allows the evolution of the curvature perturbation
to be calculated, on scales larger than the horizon scale where one
can neglect spatial gradients, using only the evolution of unperturbed
\char`\"{}separate universes\char`\"{}. The central result is that
the difference in the number of $e$-folds that occurs from different
positions on an initial flat slice of space-time to a final uniform
density slice, when compared with some fiducial value, is related
to the curvature perturbation. Writing the number of e-foldings as
a function of the initial and final time on the relevant hypersurfaces,
\begin{equation}
N\left(t,\,t_{i},\,x\right)=\int_{t_{i}}^{t}dt^{\prime}H\left(t^{\prime},\,x\right)\,,
\end{equation}
the primordial curvature perturbation can be expressed as 
\begin{equation}
\zeta\left(t,x\right)=N\left(t,\,t_{i},\,x\right)-N_{0}\left(t,\,t_{i}\right)\,,
\end{equation}
where $N_{0}\left(t,\,t_{i}\right)=\int_{t_{i}}^{t}dt^{\prime}H_{0}\left(t^{\prime}\right)$.
Taking $t_{i}=t_{*}$, the time corresponding to the modes exiting
the horizon $\left(kc_{s}=aH\right)$, the curvature perturbation
on superhorizon scales can be written in terms of partial derivatives
of $N$ with respect to the unperturbed scalar field values at horizon
exit, while holding the initial and final hypersurface constant. More
precisely 
\begin{equation}
\zeta\left(t,\,x\right)=\sum_{I}N_{,I}\left(t\right)\delta\phi_{*}^{I}(x)+\frac{1}{2}\sum_{IJ}N_{,IJ}\left(t\right)\delta\phi_{*}^{I}\left(x\right)\delta\phi_{*}^{J}\left(x\right)+\cdots\,,
\end{equation}
where $N_{,I}=\frac{\partial N}{\partial\phi_{I}^{*}}$. In momentum
space we have 
\begin{equation}
\zeta(k)=N_{,I}\delta\phi_{*}^{I}(k)+\frac{1}{2}N_{,IJ}\left[\delta\phi_{*}^{I}\star\delta\phi_{*}^{J}\right](k)+\cdots\,,\label{pertexpansion}
\end{equation}
where $\star$ indicates a convolution.

\subsection{The bispectrum}

In Fourier space the two- and three-point functions are defined, respectively,
by 
\begin{eqnarray}
\langle\zeta\left(\mathbf{k_{1}}\right)\zeta\left(\mathbf{k_{2}}\right)\rangle & = & \left(2\pi\right)^{3}\delta^{3}\left(\mathbf{k_{1}}+\mathbf{k_{2}}\right){P}_{\zeta}\left(k_{1}\right)\,,\\
\langle\zeta\left(\mathbf{k_{1}}\right)\zeta\left(\mathbf{k_{2}}\right)\zeta\left(\mathbf{k_{3}}\right)\rangle & = & \left(2\pi\right)^{3}\delta^{3}\left(\mathbf{k_{1}}+\mathbf{k_{2}}+\mathbf{k_{3}}\right)\mathcal{B}_{\zeta}\left(k_{1},k_{2},k_{3}\right)\,,
\end{eqnarray}
where $P_{\zeta}(k)$ is the power spectrum, and $B_{\zeta}\left(k_{1},k_{2},k_{3}\right)$
the bispectrum. Often the bispectrum is normalized to form the reduced
bispectrum $\fnl\left(k_{1},k_{2},k_{3}\right)$ 
\begin{equation}
B_{\zeta}\left(k_{1},k_{2},k_{3}\right)=\frac{6}{5}f_{{\rm NL}}(k_{1},k_{2},k_{3})\biggl[P_{\zeta}\left(k_{1}\right)P_{\zeta}\left(k_{2}\right)+P_{\zeta}\left(k_{2}\right)P_{\zeta}\left(k_{3}\right)+P_{\zeta}\left(k_{3}\right)P_{\zeta}\left(k_{1}\right)\biggr]\,,
\label{bispectrum}
\end{equation}

\subsection{Calculating the bispectrum with $\delta N$}

The power spectrum and bispectrum of field fluctuations at horizon
crossing follow from the two- and three-point correlations of these
perturbations as 
\begin{eqnarray}
\langle\delta\phi_{*}^{I}(\mathbf{k_{1}})\delta\phi_{*}^{J}(\mathbf{k_{2}})\rangle & = & (2\pi)^{3}G^{IJ}\frac{2\pi^{2}}{k^{3}}{\cal P}^{*}\delta\left(\mathbf{k_{1}}+\mathbf{k_{2}}\right)\\
\langle\delta\phi_{*}^{I}(\mathbf{k_{1}})\delta\phi_{*}^{J}(\mathbf{k_{2}})\delta\phi_{*}^{K}(\mathbf{k_{3}})\rangle & = & (2\pi)^{3}\frac{4\pi^{4}}{\Pi_{i}k_{i}^{3}}{\cal P^{*}}^{2}A^{IJK}(k_{1},k_{2},k_{3})\delta\left(\mathbf{k_{1}}+\mathbf{k_{2}}+\mathbf{k_{2}}\right)\,,
\end{eqnarray}
where ${\cal P}=Pk^{3}/(2\pi^{2})$. Employing the $\delta N$ expansion
one finds that 
\begin{equation}
P_{\zeta}(k)=N_{I}N_{I}P^{*}
\end{equation}
and 
\begin{equation}
f_{{\rm NL}}=f_{{\rm NL}}^{(3)}+f_{{\rm NL}}^{(4)}+\cdots\,,\label{fnl}
\end{equation}
where 
\begin{equation}
\begin{aligned}f_{{\rm NL}}^{(3)} & =\frac{5}{6}\frac{N_{,I}N_{,J}N_{,K}A^{IJK}}{\left(G^{IJ}N_{,I}N_{,J}\right)^{2}\sum_{i}k_{i}^{3}},\\
f_{{\rm NL}}^{(4)} & =\frac{5}{6}\frac{G^{IK}G^{JL}N_{,I}N_{,J}N_{,KL}}{\left(G^{IJ}N_{,I}N_{,J}\right)^{2}}\,.
\end{aligned}
\label{fnl34}
\end{equation}
Here $f_{{\rm NL}}^{(3)}$ is momentum dependent, whereas $f_{{\rm NL}}^{(4)}$
is momentum independent (which is the definition of local $\fnl$)
\footnote{Technically these results are valid only when there is not a large
hierarchy between the three wave numbers of the bispectrum and they
can all be assumed to cross the horizon at roughly the same time.
This provides a good approximation even for large hierarchies as long
as there is not a significant evolution between the horizon crossing
times of the three modes (see Refs.~\cite{Kenton:2015lxa,Kenton:2016abp}
for a full discussion)}. 
In general, the dominant contribution, $f_{{\rm NL}}^{(3)}$ or $f_{{\rm NL}}^{(4)}$,
is model dependent. For example, in the case of multiple canonical
scalar fields inflation, $f_{{\rm NL}}^{(4)}$ can become significant
. In contrast, for noncanonical models, $f_{{\rm NL}}^{(3)}$ can
become large.

For general multi-field non-canonical models in slow-roll (which is
the situation relevant to our models), utilising the In-In formalism
to calculate the statistics of the scalar field perturbations on flat
hypersurfaces at horizon crossing it was found that 
\begin{eqnarray}
P_{*}=\frac{H^{2}}{2k^{3}P_{,X}},\label{Pstar}
\end{eqnarray}
and that \cite{Gao:2008dt} 
\begin{equation}
A_{IJK}=\frac{1}{4}\sqrt{\frac{P_{,X}}{2}}\tilde{A}_{IJK},\label{aamplitude}
\end{equation}
with 
\begin{equation}
\begin{aligned}\tilde{A}^{IJK}= & G^{IJ}\epsilon^{K}\frac{u}{\epsilon}\left[\frac{4k_{1}^{2}k_{2}^{2}k_{3}^{2}}{K^{3}}-2\left(\mathbf{k_{1}}.\mathbf{k_{2}}\right)k_{3}^{2}\left(\frac{1}{K}+\frac{k_{1}+k_{2}}{K^{2}}+\frac{2k_{1}k_{2}}{K^{3}}\right)\right]\\
 & -G^{IJ}\epsilon^{K}\left[6\frac{k_{1}^{2}k_{2}^{2}}{K}+2\frac{k_{1}^{2}k_{2}^{2}\left(k_{3}+2k_{2}\right)}{K^{2}}+k_{3}k_{2}^{2}-k_{3}^{3}\right]\\
 & +G^{IJ}\left[\left(3\frac{u}{\epsilon}+4u+4\right)\tilde{\epsilon}^{K}+\tilde{\epsilon}_{,X}^{K}\frac{12H^{2}}{P_{,X}}\right]\times\\
 & \left[-\frac{k_{1}^{2}k_{2}^{2}}{K}-\frac{k_{1}^{2}k_{2}^{2}k_{3}}{K^{2}}+\left(\mathbf{k_{1}}.\mathbf{k_{2}}\right)\left(-K+\frac{\underset{i>j}{\sum}k_{i}k_{j}}{K}+\frac{k_{1}k_{2}k_{3}}{K^{2}}\right)\right]\\
 & +\frac{\epsilon^{IJ}}{\epsilon}\epsilon^{K}\left(\frac{2\lambda}{H^{2}\epsilon^{2}}-\frac{u}{\epsilon}\right)\frac{4k_{1}^{2}k_{2}^{2}k_{3}^{2}}{K^{3}}+{\rm perms}.\,,
\end{aligned}
\label{amplitude}
\end{equation}
where $K=k_{1}+k_{2}+k_{3}$, and the Hubble parameter $H$, the sound
speed squared $\left(c_{s}^{2}\right)$, and slow-roll parameters
$\left(\epsilon,\epsilon^{I},...,{\rm etc.}\right)$ are evaluated
at sound horizon exit $c_{s}k=aH$. Expressions for $c_{s}^{2}$,
$u$ and $\lambda$ are given in Ref.~\cite{Gao:2008dt} for non-Canonical
models\footnote{{We have corrected typos in the first and third lines of Eq. (\ref{amplitude})
that were present in Ref.~\cite{Gao:2008dt}.}}. In this work, we express all of these parameters in terms of three-form
quantities using Eqs.~(\ref{dual action}) and (\ref{3-formdual}).
First $u$ is defined as 
\begin{equation}
u\equiv\frac{1}{c_{s}^{2}}-1\,,\label{u}
\end{equation}
where the effective speed of sound is given by 
\begin{equation}
c_{s}^{2}=\frac{P_{,X}}{2XP_{,XX}+P_{,X}}=\frac{\underset{I}{\sum}\frac{\chi_{I}}{V_{,\chi_{I}}}}{\underset{I}{\sum}V_{,\chi_{I}\chi_{I}}^{-1}}.
\label{cs2}
\end{equation}
We also define $\lambda$, such that 
\begin{equation}
\lambda=X^{2}P_{,XX}+\frac{2}{3}X^{3}P_{,XXX}=-\sum_{I}\frac{V_{,\chi_{I}}^{3}V_{\chi_{I}\chi_{I}\chi_{I}}}{12V_{,\chi_{I}\chi_{I}}^{3}}\,.
\,\label{lambda}
\end{equation}
The various slow-roll quantities are defined by 
\begin{equation}
\epsilon\equiv-\frac{\dot{H}}{H^{2}}=\frac{3}{2}\frac{\underset{I}{\sum}\chi_{I}V_{,\chi_{I}}}{V}\left(1-\underset{I}{\sum}w_{I}^{2}\right)\,,\label{eps}
\end{equation}
\begin{equation}
\epsilon^{IJ}=\frac{P_{,X}\dot{\phi}^{I}\dot{\phi}^{J}}{2H^{2}}=\frac{P_{,X}\sqrt{X_{I}X_{J}}}{2H^{2}}=\epsilon^{I}\epsilon^{J}\,,\label{epsij}
\end{equation}
where 
\begin{equation}
\epsilon^{I}=\sqrt{\frac{X_{I}P_{,X}}{2H^{2}}}=\sqrt{\frac{3V_{,\chi_{I}}^{2}}{4V}\left(\underset{I}{\sum}\frac{\chi_{I}}{V_{,\chi_{I}}}\right)\left(1-\underset{I}{\sum}w_{I}^{2}\right)}\,,\label{epsi}
\end{equation}
\begin{equation}
\tilde{\epsilon}{}_{I}=-\frac{P_{,I}}{3\sqrt{2P_{,X}}H^{2}}=\frac{\sqrt{6}w_{I}}{3\sqrt{2\underset{I}{\sum}\frac{\chi_{I}}{V_{,\chi_{I}}}}H}\,.\label{epstilde}
\end{equation}
Using the Friedmann equation in Eq. (\ref{NFriedm-1-1}) we obtain
\begin{equation}
\begin{alignedat}{1}\tilde{\epsilon}_{,X}^{I} & =-\frac{P_{,XI}}{3\sqrt{2P_{,X}}H^{2}}+P_{,I}\left[\frac{2XP_{,XX}+P_{,X}}{9\sqrt{2P_{,X}}H^{4}}+\frac{P_{,XX}}{6\sqrt{2}P_{,X}^{3/2}H^{2}}\right]\,,\\
 & =-\sqrt{6}Hw_{I}\left[\frac{\underset{I}{\sum}V_{,\chi_{I}\chi_{I}}^{-1}}{\sqrt{2\underset{I}{\sum}\frac{\chi_{I}}{V_{,\chi_{I}}}}V}+\frac{\underset{I}{\sum}\left(V_{,\chi_{I}\chi_{I}}^{-1}V_{,\chi_{I}}^{-2}-\chi_{I}V_{,\chi_{I}}^{-3}\right)}{3\sqrt{2}\left(\underset{I}{\sum}\frac{\chi_{I}}{V_{,\chi_{I}}}\right)^{3/2}V}\right]\left(1-\underset{I}{\sum}w_{I}^{2}\right)\,.
\end{alignedat}
\label{depst}
\end{equation}
Note that the dual scalar field action in Eq. (\ref{dual action})
satisfies $P_{,XI}=0.$

In the squeezed limit i.e., $k_{2}\rightarrow0$, it can be seen from
Eq. (\ref{amplitude}) that $\fnl^{(3)}$ reduces to the order of
slow-roll parameters. Therefore $f_{{\rm NL}}^{(4)}$ is expected
to be dominant in this limit if non-Gaussianity is significant.

\subsection{The $\delta N$ for two three-forms}

\label{N derivatives}

The crucial step, when it comes to computing $f_{{\rm NL}}$, is the
calculation of the derivatives of $N$ with respect to the fields
at the sound horizon crossing. In general $N_{,I}$ and $N_{,IJ}$
evolve on superhorizon scales and except in a few cases (see e.g., Ref.~\cite{Vernizzi:2006ve})
the analytical computation of these quantities is not tractable. For
this reason we do our computations numerically using a method that
is explained in section \ref{NG23forms}.

First of all we must rewrite the derivatives in terms of three-forms.
Here we do this explicitly for two three-forms. The same procedure
can be extended trivially to $\mathbb{N}$ three-form fields. We can
infer the following relations from Eqs.~(\ref{3-formdual}) and (\ref{hubble})
relating two three-forms to the two noncanonical scalar fields 
\begin{equation}
\phi_{1}=\sqrt{6}Hw_{1}\equiv\phi_{1}\left(\chi_{1},\chi_{2},w_{1},w_{2}\right),\label{phi1}
\end{equation}
\begin{equation}
\phi_{2}=\sqrt{6}Hw_{2}\equiv\phi_{2}\left(\chi_{1},\chi_{2},w_{1},w_{2}\right),\label{phi1-1}
\end{equation}
It is highly nontrivial to invert the relations in Eqs. (\ref{phi1})
and (\ref{phi1-1}). 
While the fields are slowly rolling, one can verify that the approximation
$w_{I}\approx\sqrt{\frac{3}{2}}\chi_{I}$ is accurately satisfied
(see Ref.\cite{Kumar:2014oka}). As a consequence, we express the
$N$ derivatives $N_{,I}$ and $N_{,IJ}$ in terms of the two three-forms
$\chi_{1},\,\chi_{2}$ as 
\begin{equation}
\frac{\partial N}{\partial\phi_{1}^{*}}=\frac{\partial N}{\partial\chi_{1}^{*}}\frac{\partial\chi_{1}^{*}}{\partial\phi_{1}^{*}}+\frac{\partial N}{\partial\chi_{2}^{*}}\frac{\partial\chi_{2}^{*}}{\partial\phi_{1}^{*}}\,,\label{firstderivatives}
\end{equation}
\begin{equation}
\begin{aligned}\frac{\partial^{2}N}{\partial\phi_{1}^{*}\partial\phi_{2}^{*}}= & \frac{\partial N}{\partial\chi_{1}^{*}}\frac{\partial^{2}\chi_{1}^{*}}{\partial\phi_{1}^{*}\partial\phi_{2}^{*}}+\frac{\partial N}{\partial\chi_{2}^{*}}\frac{\partial^{2}\chi_{2}^{*}}{\partial\phi_{1}^{*}\partial\phi_{2}^{*}}+\frac{\partial^{2}N}{\partial\chi_{1}^{*2}}\frac{\partial\chi_{1}^{*}}{\partial\phi_{1}^{*}}\frac{\partial\chi_{1}^{*}}{\partial\phi_{2}^{*}}\\
 & +\frac{\partial^{2}N}{\partial\chi_{2}^{*2}}\frac{\partial\chi_{2}^{*}}{\partial\phi_{1}^{*}}\frac{\partial\chi_{2}^{*}}{\partial\phi_{2}^{*}}+\frac{\partial^{2}N}{\partial\chi_{1}^{*}\partial\chi_{2}^{*}}\frac{\partial\chi_{1}^{*}}{\partial\phi_{1}^{*}}\frac{\partial\chi_{2}^{*}}{\partial\phi_{2}^{*}}+\frac{\partial^{2}N}{\partial\chi_{1}^{*}\partial\chi_{2}^{*}}\frac{\partial\chi_{1}^{*}}{\partial\phi_{2}^{*}}\frac{\partial\chi_{2}^{*}}{\partial\phi_{1}^{*}}\,,
\end{aligned}
\label{d2np12-1}
\end{equation}
\begin{equation}
\frac{\partial^{2}N}{\partial\phi_{1}^{*2}}=\frac{\partial N}{\partial\chi_{1}^{*}}\frac{\partial^{2}\chi_{1}^{*}}{\partial\phi_{1}^{*2}}+\frac{\partial N}{\partial\chi_{2}^{*}}\frac{\partial^{2}\chi_{2}^{*}}{\partial\phi_{1}^{*2}}+\frac{\partial^{2}N}{\partial\chi_{1}^{*2}}\left(\frac{\partial\chi_{1}^{*}}{\partial\phi_{1}^{*}}\right)^{2}+\frac{\partial^{2}N}{\partial\chi_{2}^{*2}}\left(\frac{\partial\chi_{2}^{*}}{\partial\phi_{1}^{*}}\right)^{2}+2\frac{\partial^{2}N}{\partial\chi_{1}^{*}\partial\chi_{2}^{*}}\frac{\partial\chi_{1}^{*}}{\partial\phi_{1}^{*}}\frac{\partial\chi_{2}^{*}}{\partial\phi_{1}^{*}}\,.\label{d2np12}
\end{equation}
derivatives of $\phi_{2}$. These equations define the relations among
the $N$ derivatives ($N_{,I}$ and $N_{,IJ}$) with respect to scalar
field $\phi_{I}^{*}$ to the $N$ derivatives with respect to three-form
fields at horizon crossing $\frac{\partial N}{\partial\chi_{1}^{*}}\,,\,\frac{\partial N}{\partial\chi_{2}^{*}}\,,\,\frac{\partial^{2}N}{\partial\chi_{1}^{*}\partial\chi_{2}^{*}}\,,\,\frac{\partial^{2}N}{\partial\chi_{1}^{*2}}\,,\,\frac{\partial^{2}N}{\partial\chi_{2}^{*2}}$.
In other words, we have indirectly transported the $\delta N$ formalism
from scalar fields to three-form fields. However, we still need to
calculate the derivatives of the three-form fields with respect to
the dual scalar fields. For this purpose we differentiate the relations
(\ref{phi1}) and (\ref{phi1-1}) keeping in mind that $\phi_{1}$
and $\phi_{2}$ are independent fields. Then we have that 
\begin{equation}
\frac{d\phi_{1}}{d\phi_{1}}=\frac{1}{\sqrt{6}w_{1}}\frac{\partial H}{\partial\phi_{1}}+\frac{1}{\sqrt{6}H}\frac{\partial w_{1}}{\partial\phi_{1}}=1\,.\label{dph11}
\end{equation}
\begin{equation}
\frac{d\phi_{1}}{d\phi_{2}}=\frac{1}{\sqrt{6}w_{1}}\frac{\partial H}{\partial\phi_{2}}+\frac{1}{\sqrt{6}H}\frac{\partial w_{1}}{\partial\phi_{2}}=0\,.\label{dph11-1}
\end{equation}
\begin{equation}
\frac{d\phi_{2}}{d\phi_{1}}=\frac{1}{\sqrt{6}w_{2}}\frac{\partial H}{\partial\phi_{2}}+\frac{1}{\sqrt{6}H}\frac{\partial w_{2}}{\partial\phi_{2}}=1\,.\label{dph11-1-1}
\end{equation}
\begin{equation}
\frac{d\phi_{2}}{d\phi_{2}}=\frac{1}{\sqrt{6}w_{2}}\frac{\partial H}{\partial\phi_{1}}+\frac{1}{\sqrt{6}H}\frac{\partial w_{2}}{\partial\phi_{1}}=0\,.\label{dph11-1-1-1}
\end{equation}
Solving Eqs.~(\ref{dph11})-(\ref{dph11-1-1-1}) for a potential
of the form $V=V\left(\chi_{1}\right)+V\left(\chi_{2}\right)$, we
obtain 
\begin{equation}
\begin{aligned}\frac{\partial\chi_{1}}{\partial\phi_{1}}= & \frac{\chi_{2}V_{,\chi_{2}}+H^{2}\left(6-9\chi_{1}^{2}\right)}{3H\left(6H^{2}+\chi_{1}V_{,\chi_{1}}+\chi_{2}V_{,\chi_{2}}\right)}\\
\frac{\partial\chi_{1}}{\partial\phi_{2}}= & -\frac{\chi_{1}\left(V_{,\chi_{2}}+9H^{2}\chi_{2}\right)}{3H\left(6H^{2}+\chi_{1}V_{,\chi_{1}}+\chi_{2}V_{,\chi_{2}}\right)}
\end{aligned}
\label{Qf}
\end{equation}
\begin{equation}
\begin{aligned}\frac{\partial^{2}\chi_{1}}{\partial\phi_{1}^{2}}= & \frac{-1}{9H^{2}\left(6H^{2}+\chi_{1}V_{,\chi_{1}}+\chi_{2}V_{,\chi_{2}}\right)^{3}}\lbrace\chi_{1}V_{,\chi_{1}}^{2}\left[\chi_{2}\left(\chi_{2}V_{,\chi_{2}\chi_{2}}+2V_{,\chi_{2}}\right)+H^{2}\left(9\chi_{1}^{2}-6\right)\right]\\
 & -2V_{,\chi_{1}}\left[-3H^{2}\chi_{2}\left(3V_{,\chi_{2}\chi_{2}}\chi_{1}^{2}\chi_{2}+6V_{,\chi_{2}}\chi_{1}^{2}+4V_{,\chi_{2}}\right)-V_{,\chi_{2}}^{2}\chi_{2}^{2}+18H^{4}\left(3\chi_{1}^{2}-2\right)\right]\\
 & +\chi_{1}V_{,\chi_{1}\chi_{1}}\left(\chi_{2}V_{,\chi_{2}}+H^{2}\left(6-9\chi_{1}^{2}\right)\right)^{2}\\
 & -9\chi_{1}H^{2}\left(-3H^{2}\chi_{2}\left(3V_{,\chi_{2}\chi_{2}}\chi_{1}^{2}\chi_{2}+12V_{,\chi_{2}}\right)-3V_{,\chi_{2}}^{2}\chi_{2}^{2}+54H^{4}\left(3\chi_{1}^{2}-2\right)\right)\rbrace\,.
\end{aligned}
\label{Q1F1F1}
\end{equation}
\begin{equation}
\begin{aligned}\frac{\partial^{2}\chi_{1}}{\partial\phi_{2}^{2}}= & \frac{-1}{9H^{2}\left(6H^{2}+\chi_{1}V_{,\chi_{1}}+\chi_{2}V_{,\chi_{2}}\right)^{3}}\lbrace\chi_{1}\left[18V_{,\chi_{2}}H^{2}\chi_{2}\left(V_{,\chi_{1}\chi_{1}}\chi_{1}^{2}-18H^{2}\right)-2V_{,\chi_{2}}^{3}\chi_{2}\right]\\
 & +\chi_{1}V_{,\chi_{2}}^{2}\left[\chi_{1}\left(\chi_{1}V_{,\chi_{1}\chi_{1}}-2V_{,\chi_{1}}\right)-3H^{2}\left(3\chi_{2}^{2}+10\right)\right]+\chi_{1}V_{,\chi_{2}\chi_{2}}\left[V_{,\chi_{1}}\chi_{1}+H^{2}\left(6-9\chi_{2}^{2}\right)\right]^{2}\\
 & +9\chi_{1}H^{2}\left[3H^{2}\chi_{1}\left(3V_{,\chi_{1}\chi_{1}}\chi_{1}\chi_{2}^{2}+4V_{,\chi_{1}}\right)+V_{,\chi_{1}}^{2}\chi_{1}^{2}-18H^{4}\left(9\chi_{2}^{2}-2\right)\right]\rbrace\,.
\end{aligned}
\label{q1f2f2}
\end{equation}
\begin{equation}
\begin{aligned}\frac{\partial^{2}\chi_{1}}{\partial\phi_{1}\partial\phi_{2}}= & \frac{1}{9H^{2}\left(6H^{2}+\chi_{1}V_{,\chi_{1}}+\chi_{2}V_{,\chi_{2}}\right)^{3}}\lbrace-V_{,\chi_{2}}^{3}\chi_{2}^{2}+V_{,\chi_{2}}^{2}\chi_{2}\left[V_{,\chi_{1}\chi_{1}}\chi_{1}^{2}+3H^{2}\left(-4+3\chi_{1}^{2}-3\chi_{2}^{2}\right)\right]\\
 & +V_{,\chi_{2}}\left[3H^{2}\chi_{1}\left(V_{,\chi_{1}\chi_{1}}\chi_{1}\left(-3\chi_{1}^{2}+3\chi_{2}^{2}+2\right)+3V_{,\chi_{1}}\left(\chi_{1}^{2}-\chi_{2}^{2}+2\right)\right)+V_{,\chi_{1}}^{2}\chi_{1}^{2}\right]\\
 & +36V_{,\chi_{2}}H^{4}\left(6\chi_{1}^{2}-3\chi_{2}^{2}-1\right)+\text{\ensuremath{\chi_{2}}}\left(V_{,\chi_{2}\chi_{2}}V_{,\chi_{1}}^{2}\chi_{1}^{2}+3V_{,\chi_{2}\chi_{2}}V_{,\chi_{1}}H^{2}\chi_{1}\left(3\chi_{1}^{2}-3\chi_{2}^{2}+2\right)\right)+\\
 & \chi_{2}\left[162H^{6}\left(9\chi_{1}^{2}-2\right)+27H^{4}\chi_{1}\left(\chi_{1}\left(-3V_{,\chi_{1}\chi_{1}}\chi_{1}^{2}+2V_{,\chi_{1}\chi_{1}}-3V_{,\chi_{2}\chi_{2}}\chi_{2}^{2}+2V_{,\chi_{2}\chi_{2}}\right)+4V_{,\chi_{1}}\right)\right]\rbrace\,.
\end{aligned}
\label{q1f1f2}
\end{equation}
The remaining derivatives can be obtained from these by interchanging
$1\leftrightarrow2$. Following Eqs.~(\ref{firstderivatives})-(\ref{d2np12})
the quantities obtained in Eqs.~(\ref{Qf})-(\ref{q1f1f2}) 
are to be evaluated at $kc_{s}=aH$. However, the derivatives of $N$
with respect to the three-form fields evolve on superhorizon scales.

\section{Two three-form non-Gaussianity and observational data}

\label{NG23forms}

\begin{figure}[h!]
\includegraphics[width=3.5in]{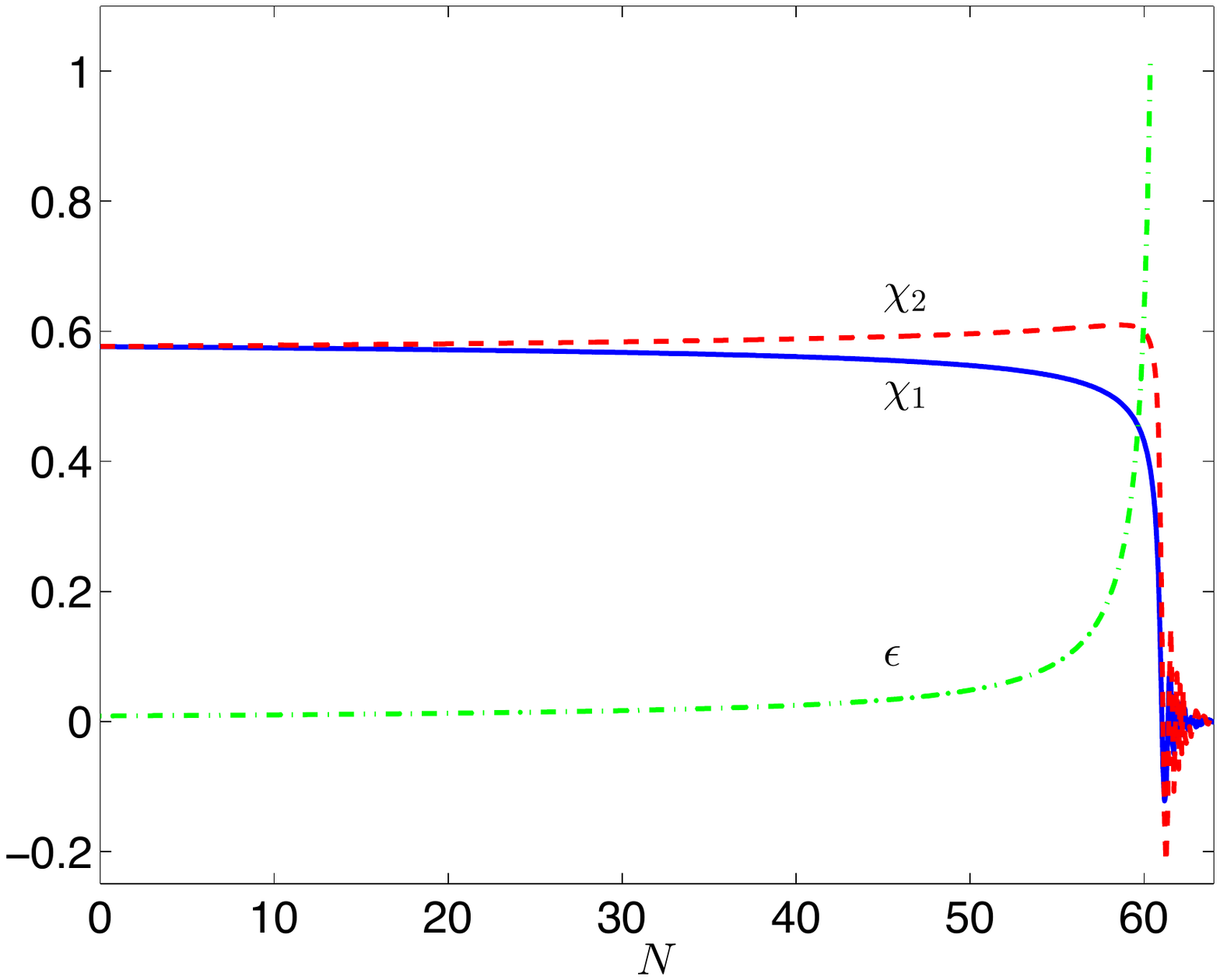} \caption{\label{23formdynamics} The numerical solutions of (\ref{NDiff-syst-1-1-3})
for $\chi_{1}\left(N\right)$ (solid line) and $\chi_{2}\left(N\right)$
(dashed line). The dash-dotted line corresponds to the slow-roll parameter
$\epsilon\left(N\right)$ and $\epsilon=1$ indicates the end of inflation
at $N=60.35$. We have considered the potentials $V_{1}=V_{10}\left(\chi_{1}^{2}+b_{1}\chi_{1}^{4}\right)$
and $V_{2}=V_{20}\left(\chi_{2}^{2}+b_{2}\chi_{2}^{4}\right)$ with
$V_{10}=1,\:V_{20}=0.93,\:b_{1,2}=-0.35$ and taken the initial conditions
$\chi_{1}\left(0\right)\approx0.5763,\,\chi_{2}\left(0\right)\approx0.5766,\,\chi_{1}^{\prime}\left(0\right)=-0.000224,\,\chi_{2}^{\prime}\left(0\right)=0.00014$.}
\end{figure}

In this section, we aim to update the observational status of two
three-form inflation \cite{Kumar:2014oka} by means of calculating
the reduced bispectrum $f_{{\rm NL}}$. {Assuming a slow-roll regime, from Eq.
(\ref{eps}) and in the case of two three-forms, $\epsilon\ll1$ leads to
the following condition 
\begin{equation}
w_{1}^{2}+w_{2}^{2}\approx1\,,\label{23formC1}
\end{equation}
which we can parametrize as
\begin{equation}
\begin{aligned}w_{1}\approx\:\cos\theta\,,\\
w_{2}\approx\:\sin\theta\,.
\end{aligned}
\label{a-value}
\end{equation}
Subsequently, from (\ref{Nauton-n-n}), we can establish the initial
conditions for the field derivatives as
\begin{eqnarray}
\begin{cases}
\chi_{1}^{\prime}\approx3\left(\sqrt{\frac{2}{3}}\:\cos\theta-\chi_{1}\right),\\
\chi_{2}^{\prime}\approx3\left(\sqrt{\frac{2}{3}}\;\sin\theta-\chi_{2}\right).
\end{cases}\label{x1x2plateau}
\end{eqnarray}
}

{As described in Sec. \ref{N3formmodel}, there exist two kinds of inflationary dynamics
to consider, namely, type I and type II solutions. In the type
I case, the trajectories in field space are straight lines and
the two three-form fields stay near the fixed points
\begin{equation}
\chi_{1c}=\sqrt{\frac{2}{3}}\cos\theta_{c}\quad,\quad\chi_{2c}=\sqrt{\frac{2}{3}}\sin\theta_{c}\quad,\quad\theta_{c}=\arctan\left(\frac{\lambda_{2}}{\lambda_{1}}\right)\Big|_{\chi_{1}=\chi_{1c},\chi_{2}=\chi_{2c}}\,.\label{IctypeI}
\end{equation}
Notice that we have used Eqs. (\ref{fixed-P2}) and (\ref{a-value}) to obtain (\ref{IctypeI}). Given the potential $V\left(\chi_{1},\,\chi_{2}\right)$
we can find the critical angle $\theta_{c}$} {that
gives us initial conditions for which the three-form fields evolve almost
identically and generate straight line trajectories in field space.
In this scenario, there are no isocurvature perturbations produced
during inflation and as a consequence the reduced local bispectrum $f_{NL}$
is negligible. In the type II case,
we choose an initial condition away from $\theta_{c}$ that leads
to a situation where three-form fields evolve away from $\left(\chi_{1c},\,\chi_{2c}\right)$
leading to curved trajectories in field space. Moreover, different types of potentials
$V\left(\chi_{1},\,\chi_{2}\right)$ will diversely affect the particular form of the curved trajectory. 
In Refs. \cite{Felice-1,felice2} suitable potentials for three-forms were proposed 
for being adequate to avoid ghost and Laplacian instabilities ($0<c_{s}^{2}\lesssim1$). It was
shown that the potentials with a quadratic behavior when $\chi_{I}\to0$
were free from ghost instabilities and displayed 
an oscillatory behavior near the end of inflation \cite{Kumar:2014oka,Barros:2015evi}.
In this regard, potentials of the form $V\left(\chi_{I}\right)=a\chi_{I}^{2}+b\chi_{I}^{2n}$
with $b<0$ are free from ghost instabilities and consistent with
a sound speed $0<c_{s}^{2}\lesssim1$. 
Finally, it is important to point out that we can also have other more generalized
potentials like $V\left(\chi_{I}\right)=\exp\left(\nu\chi_{I}^{2}\right)-1$
or $\tanh\left(\nu\chi_{I}^{2}\right)$ \cite{Felice-1,Kumar:2014oka}. }

In Ref. \cite{Kumar:2014oka}, type II solutions with {potentials
$V\left(\chi_{1},\,\chi_{2}\right)=V_{10}f\left(\chi_{1}\right)+V_{20}f\left(\chi_{2}\right)$,
where $f\left(\chi_{I}\right)=\chi_{I}^{2}+b\chi_{I}^{2n}$, using}
tuned initial conditions were shown to be consistent with Planck 2013
data, predicting the scalar spectral index $n_{s}\sim0.967$ and the tensor
to scalar ratio $r\sim0.0422$. {The fine-tuning process 
consists in introducing a tiny asymmetry in the potential
by means of taking $V_{10}\neq V_{20}$. This asymmetry corresponds
to a curved trajectory in field space, thus giving rise to a
controlled growth of curvature perturbation on superhorizon scales.
This fine-tuning is essential to keep} 
the running of the spectral index $\left(\frac{dn_{s}}{dlnk}\right)$
negligible and compatible
with the observational data.\footnote{{See the discussion in section 5 of Ref. \cite{Kumar:2014oka}
for details concerning the effect of isocurvature modes on the running spectral
index.}}. The {two three-forms }dynamics that give rise to
these consistent predictions are plotted in Fig.~\ref{23formdynamics}.
We have taken the same initial conditions and the parameter values
\footnote{The initial conditions considered in Fig.~\ref{23formdynamics} correspond
to the values of three-form fields at horizon crossing, whereas initial
conditions in Ref. \cite{Kumar:2014oka} were taken at an instant preceding the slow-roll regime. Nevertheless, we study the same inflationary trajectory which
was proved to be compatible with the Planck 2013 data in Ref. \cite{Kumar:2014oka}. } as in \cite{Kumar:2014oka}.

The observational prediction of non-Gaussianity for multifield inflation
is deeply associated with the evolution of isocurvature perturbations.
In the single field inflation the statistics of the curvature pertrubation
evaluated at horizon exit can be confronted with the observation.
This is because the curvature perturbation is conserved on superhorizon
scales if the system is adiabatic \cite{Lyth2004,Rigopoulos:2003ak,Christopherson:2008ry}.
Whereas for multifield models, the statistics evolve on superhorizon
scales and non-Gaussianity can be generated as a consequence of the
presence of isocurvature perturbations. This can happen in two regimes,
namely, (i) during inflation \cite{Byrnes:2008wi,Peterson:2010np,Elliston:2011dr,Elliston:2012wm} and
(ii) after inflation such as in the curvaton model \cite{Mollerach:1989hu,Linde:1996gt,Enqvist:2001zp,Lyth:2001nq,Moroi:2001ct,Enqvist:2005pg,Linde:2005yw,Malik:2006pm,Sasaki:2006kq,Meyers:2013gua,Elliston:2014zea,Byrnes:2014xua}.
In general the statistics continue to evolve until all isocurvature
perturbations decay, the so-called adiabatic limit \cite{Elliston:2011dr}.
We evaluate $f_{{\rm NL}}$ at the end of inflation, this is
a good approximation as long as reheating proceeds quickly, and curvaton
type effects do not occur. 
\begin{figure}[h!]
\includegraphics[width=4in]{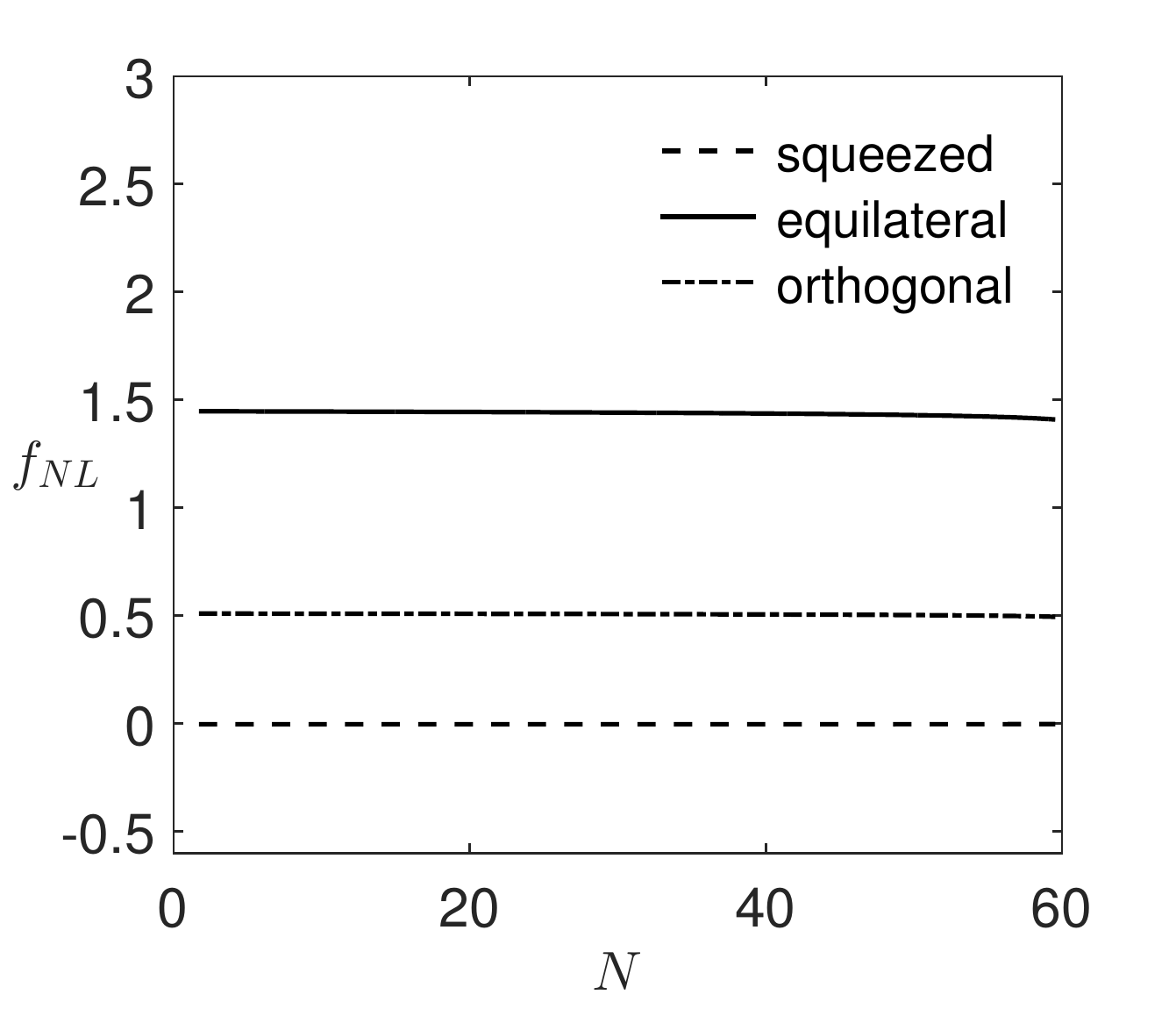}\caption{\label{fNLshape} In this plot we depict $f_{{\rm NL}}$ against $N$
for squeezed $\left(k_{2}\ll k_{1}=k_{3}\right)$ equilateral $\left(k_{1}=k_{2}=k_{3}\right)$
and orthogonal $\left(k_{1}=2k_{2}=2k_{3}\right)$ configurations.
We have considered the potentials $V_{1}=V_{10}\left(\chi_{1}^{2}+b_{1}\chi_{1}^{4}\right)$
and $V_{2}=V_{20}\left(\chi_{2}^{2}+b_{2}\chi_{2}^{4}\right)$ with
$V_{10}=1,\:V_{20}=0.93,\:b_{1,2}=-0.35$ and taken the initial conditions
$\chi_{1}\left(0\right)\approx0.5763,\,\chi_{2}\left(0\right)\approx0.5766,\,\chi_{1}^{\prime}\left(0\right)=-0.000224,\,\chi_{2}^{\prime}\left(0\right)=0.00014$.}
\end{figure}

To calculate $f_{{\rm NL}}$ given in Eq. (\ref{fnl}), we need to
compute the $N$ derivatives with respect to the initial conditions
of three-form fields defined in Eqs. (\ref{firstderivatives})-(\ref{d2np12}).
To compute these numerically, we define the following discrete derivatives
that can in principle, be extended to any number of fields, 
\begin{equation}
\begin{aligned}N_{,\chi_{1}^{*}} & =\frac{N\left(\chi_{1}^{*}+\Delta\chi_{1}\,,\,\chi_{2}^{*}\right)-N\left(\chi_{1}^{*}-\Delta\chi_{1}\,,\,\chi_{2}^{*}\right)}{2\Delta\chi_{1}},\\
N_{,\chi_{1}^{*}\chi_{1}^{*}} & =\frac{N\left(\chi_{1}^{*}+\Delta\chi_{1}\,,\,\chi_{2}^{*}\right)-2N\left(\chi_{1}^{*}\right)+N\left(\chi_{1}^{*}+\Delta\chi_{1}\,,\,\chi_{2}^{*}\right)}{\Delta\chi_{1}^{2}},\\
N_{,\chi_{1}^{*}\chi_{2}^{*}} & =\left[N\left(\chi_{1}^{*}+\Delta\chi_{1}\,,\,\chi_{2}^{*}+\Delta\chi_{2}\right)-N\left(\chi_{1}^{*}+\Delta\chi_{1}\,,\,\chi_{2}^{*}-\Delta\chi_{2}\right)-\right.\\
 & ~~~\left.N\left(\chi_{1}^{*}-\Delta\chi_{1}\,,\,\chi_{2}^{*}+\Delta\chi_{2}\right)+N\left(\chi_{1}^{*}-\Delta\chi_{1}\,,\,\chi_{2}^{*}-\Delta\chi_{2}\right)\right](4\Delta\chi_{1}^{2})^{-1},
\end{aligned}
\label{discretederivatives}
\end{equation}
and similarly we can obtain the remaining derivatives by interchanging
$1\leftrightarrow2$. In the above expression, $N\left(\chi_{1},\chi_{2}\right)$
is the number of $e$-foldings that occur starting at initial conditions
$\{\chi_{1}^{*},\chi_{2}^{*}\}$ and ending at a given final energy
density. This final energy density is defined by the condition that
$N\left(\chi_{1},\chi_{2}\right)=60.35$ at the point $\epsilon=1$.
That is the central point in the finite difference represents a trajectory
that undergoes $60$ $e$-folds of inflation, from the initial field
value until inflation ends, and the density at that time is used as
the final density for all the other points in the difference scheme.
These other points therefore represent slightly different amounts
of inflation, and we note that their associated trajectories do
not end exactly at the point $\epsilon=1$. In our numerical results
we take $\Delta\chi_{I}\sim10^{-5}$. Using the $N$ derivatives calculated
from (\ref{discretederivatives}) and evaluating the amplitude given
by Eq. (\ref{aamplitude}), we compute $f_{{\rm NL}}$ in (\ref{fnl}).
We obtain the momentum independent contribution $f_{NL}^{(4)}$ in
(\ref{fnl34}) to be very small $\mathcal{O}\left(10^{-3}\right)$.
In Fig. \ref{fNLshape} we plot the total $f_{{\rm NL}}$ versus $N$
for squeezed ($k_{2}\ll k_{1}=k_{3}$), equilateral ($k_{1}=k_{2}=k_{3}$)
and orthogonal ($k_{1}=2k_{2}=2k_{3}$) triangles.

\begin{figure}[h!]
\centering\includegraphics[width=4in]{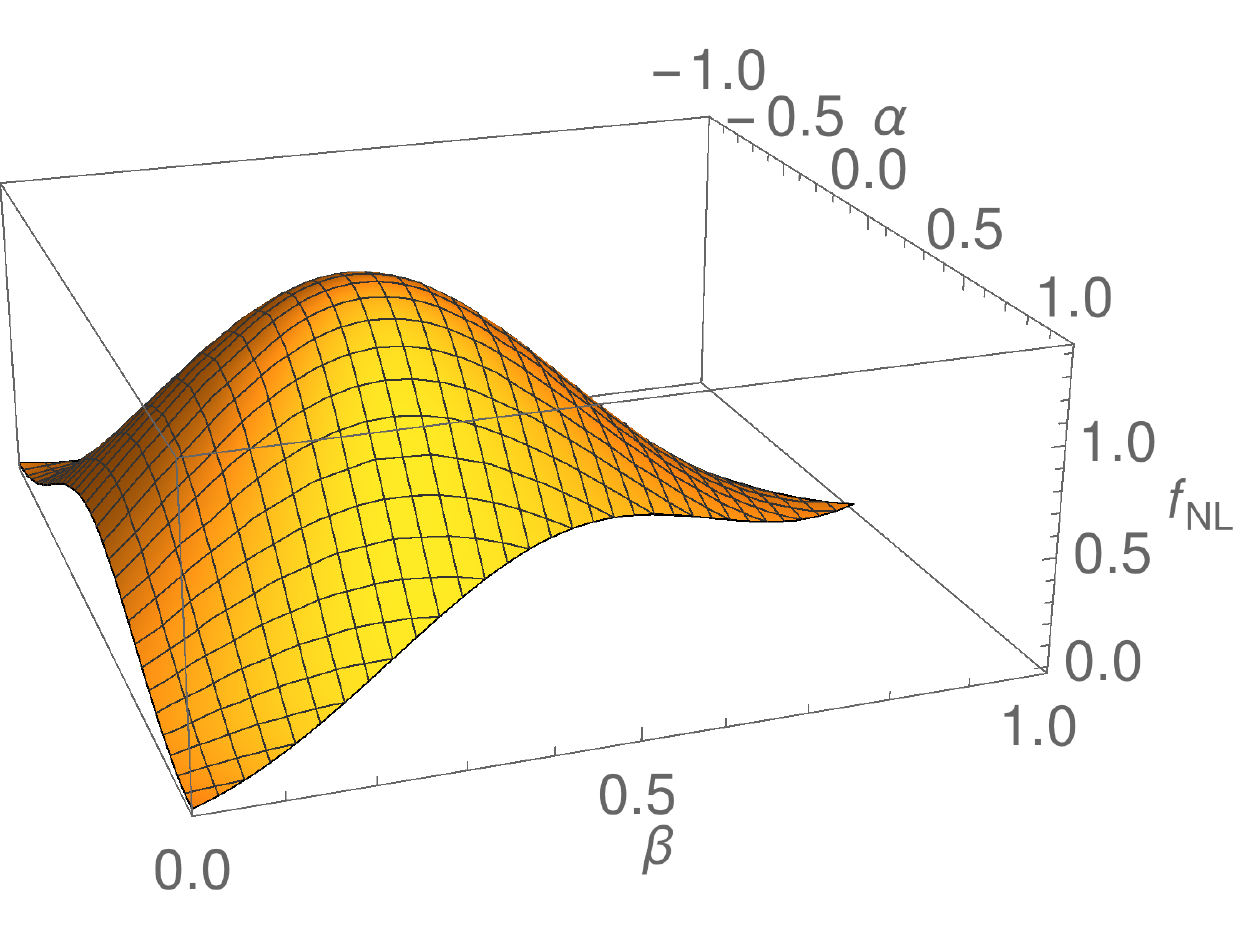} \caption{\label{fig2} Graphical representation of the non-Gaussianity shape $f_{{\rm NL}}\left(\alpha,\,\beta\right)$.
We have considered the potentials $V_{1}=V_{10}\left(\chi_{1}^{2}+b_{1}\chi_{1}^{4}\right)$
and $V_{2}=V_{20}\left(\chi_{2}^{2}+b_{2}\chi_{2}^{4}\right)$ with
$V_{10}=1,\:V_{20}=0.93,\:b_{1,2}=-0.35$ and taken the initial conditions
$\chi_{1}\left(0\right)\approx0.5763,\,\chi_{2}\left(0\right)\approx0.5766,\,\chi_{1}^{\prime}\left(0\right)=-0.000224,\,\chi_{2}^{\prime}\left(0\right)=0.00014$. }
\end{figure}

It is convenient to express the reduced bispectrum in terms of the
following independent variables \cite{Fergusson:2008ra,Fergusson:2009nv}
\begin{equation}
\alpha=\frac{k_{2}-k_{3}}{k}\quad,\quad\beta=\frac{k-k_{1}}{k}\quad\text{where}\quad k=\frac{k_{1}+k_{2}+k_{3}}{2}\,,
\end{equation}
where $0\leq\beta\leq1$ and, $-\left(1-\beta\right)\leq\alpha\leq\left(1-\beta\right)$.
In Fig.~\ref{fig2} we depict the shape of a slice through the reduced
bispectrum $f_{{\rm NL}}\left(k_{1},k_{2},k_{3}\right)$ at $N=60$
using these variables. The bispectrum shape reveals details about
the dominant interaction contributions \cite{Babich:2004gb}. In general,
the presence of a signal in the squeezed limit represents the interaction
of the long wavelength mode, which already exited the horizon, with
the short wavelength modes still being within the horizon. This can
happen in the case where more than one light scalar field drives the
period of inflation. When, instead, we observe a peak in the equilateral
limit, the dominant interaction between the fields occurs when the
modes are exiting the horizon at the same time during inflation. This
is taken to be the distinctive feature of models with a noncanonical
kinetic term or models involving higher derivative interactions
\cite{Chen:2010xka}. In the case of multiple noncanonical scalar
field inflation (which is effectively happening in the two three-form
inflation scenario), it is possible that we would encounter a mixture
of shapes \cite{Babich:2004gb,Chen:2010xka}. Although in the example
we explored there is no significant signal in the squeezed limit.

\section{Conclusions}

\label{Concl}

In this article we presented a generic framework to compute primordial
non-Gaussianity in the case of multiple three-form field inflation.
We followed the $\delta N$ formalism which is a well-known method
to study the evolution of curvature perturbations on superhorizon
scales in the case of multiple scalar fields. Because of the fact that
the three-form fields are dual to noncanonical scalar fields, which
was shown in \cite{Mulryne:2012ax}, we developed an indirect methodology
to implement $\delta N$ formalism to three-form fields. For a specific
case of two three-form fields, we derived a relation between the derivatives
of $N$ with respect to unperturbed values of scalar field duals at
horizon exit $c_{s}k=aH$ and the $N$ derivatives with respect to
three-form fields. We employed a numerical finite difference approach
for this purpose. We computed the bispectrum at horizon exit for the
two three-form field case using known expressions for three-point
field space correlations for a general multiscalar field model. Then
using the $N$ derivatives we determined the complete superhorizon
evolution of $f_{{\rm NL}}$ for squeezed, equilateral and orthogonal
configurations until the end of inflation. We considered a suitable
choice of potentials and specific values of model parameters that
were consistent with $n_{s}\sim0.967$ and $r\sim0.0422$ \cite{Kumar:2014oka}.
We obtained the corresponding $f_{{\rm NL}}$ predictions for the
two three-form inflationary model as $f_{{\rm NL}}^{{\rm sq}}\sim-2.6\times10^{-3},\,f_{{\rm NL}}^{{\rm eq}}\sim1.409,\,f_{{\rm NL}}^{{\rm orth}}\sim0.495$.
Therefore, the model is well within the observational bounds of Planck
2015 data but in principal, could be falsifiable with the future probes.

{We have computed $f_{{\rm NL}}$ for two three-forms with potentials
of the form $\chi_{I}+b\chi_{I}^{4}$, but our results may not be
significantly different with more generic potentials like $\exp\left(\nu\chi_{I}^{2}\right)-1$
or $\tanh\left(\nu\chi_{I}^{2}\right)$, under an appropriate fine-tuning in the initial conditions. 
From the conclusions drawn from the two three-form scenario (which is simpler), we cannot precisely anticipate the 
generation of non-Gaussianities in $\mathbb{N}$ three-form inflation beyond the fact
that the existence of curved trajectories in field space is also expected in the more complex case. 
Therefore, one can extend the present work to $\mathbb{N}$ three-form fields but
in such cases a much more careful analysis is needed in fine tuning
the parameters and initial conditions such that $\left(n_{s},\,r\right)$
and more importantly the running of spectral index $\left(\frac{dn_{s}}{dlnk}\right)$
are well within the current observational bounds. Another interesting
possibility, to extend this study, is to explore a curvaton type of scenario
with three-form fields with an adequate choice of potentials. Finally, the  study of the
trispectrum in this model constitutes an interesting direction that we consider for future investigation.}

\acknowledgments K.S.K. is grateful to the Instituto de Astrof\'{i}sica
e Ci\^encias do Espa\c{c}o, Universidade de Lisboa for the hospitality where
part of this work was done. K.S.K. is supported by the FCT PhD grant
SFRH/BD/51980/2012 from the portuguese agency Funda\c{c}\~ao para a Ci\^encia e technologia. D.J.M is supported by a Royal Society University
Research Fellowship. N.J.N thanks Queen Mary University of London
for hospitality. This research work is supported by the grants UID/MAT/00212/2013
and UID/FIS/04434/2013. 
{We would like to thank the anonymous referee who helped us to improve the clarity of this paper.}

\bibliographystyle{utphys}
\bibliography{References}

\providecommand{\href}[2]{#2}\begingroup\raggedright\begin{thebibliography}{10}

\bibitem{Guth:1980zm}
A.~H. Guth, ``{The Inflationary Universe: A Possible Solution to the Horizon
  and Flatness Problems},''
\href{http://dx.doi.org/10.1103/PhysRevD.23.347}{{\em Phys.Rev.} {\bfseries
  D23} (1981) 347--356}.

\bibitem{Ade:2015xua}
{\bfseries Planck} Collaboration, P.~A.~R. Ade {\em et~al.}, ``{Planck 2015
  results. XIII. Cosmological parameters},''
\href{http://arxiv.org/abs/1502.01589}{{\ttfamily arXiv:1502.01589
  [astro-ph.CO]}}.

\bibitem{Ade:2015lrj}
{\bfseries Planck} Collaboration, P.~A.~R. Ade {\em et~al.}, ``{Planck 2015
  results. XX. Constraints on inflation},''
\href{http://arxiv.org/abs/1502.02114}{{\ttfamily arXiv:1502.02114
  [astro-ph.CO]}}.

\bibitem{Ade:2015ava}
{\bfseries Planck} Collaboration, P.~A.~R. Ade {\em et~al.}, ``{Planck 2015
  results. XVII. Constraints on primordial non-Gaussianity},''
\href{http://arxiv.org/abs/1502.01592}{{\ttfamily arXiv:1502.01592
  [astro-ph.CO]}}.

\bibitem{Ade:2015tva}
{\bfseries BICEP2, Planck} Collaboration, P.~Ade {\em et~al.}, ``{Joint
  Analysis of BICEP2/Keck Array and Planck Data},''
  \href{http://dx.doi.org/10.1103/PhysRevLett.114.101301}{{\em Phys. Rev.
  Lett.} {\bfseries 114} (2015) 101301},
\href{http://arxiv.org/abs/1502.00612}{{\ttfamily arXiv:1502.00612
  [astro-ph.CO]}}.

\bibitem{Martin:2014vha}
J.~Martin, C.~Ringeval, and V.~Vennin, ``{Encyclopaedia Inflationaris},''
  \href{http://dx.doi.org/10.1016/j.dark.2014.01.003}{{\em Phys.Dark Univ.}
  (2014) },
\href{http://arxiv.org/abs/1303.3787}{{\ttfamily arXiv:1303.3787
  [astro-ph.CO]}}.

\bibitem{Martin:2015dha}
J.~Martin, ``{The Observational Status of Cosmic Inflation after Planck},''
\newblock 2015.
\newblock
\href{http://arxiv.org/abs/1502.05733}{{\ttfamily arXiv:1502.05733
  [astro-ph.CO]}}.
\newblock

\bibitem{Tsujikawa:2014rta}
S.~Tsujikawa, ``{Distinguishing between inflationary models from cosmic
  microwave background},'' \href{http://dx.doi.org/10.1093/ptep/ptu047}{{\em
  PTEP} {\bfseries 2014} no.~6, (2014) 06B104},
\href{http://arxiv.org/abs/1401.4688}{{\ttfamily arXiv:1401.4688
  [astro-ph.CO]}}.

\bibitem{Maldacena:2002vr}
J.~M. Maldacena, ``{Non-Gaussian features of primordial fluctuations in single
  field inflationary models},''
  \href{http://dx.doi.org/10.1088/1126-6708/2003/05/013}{{\em JHEP} {\bfseries
  0305} (2003) 013},
\href{http://arxiv.org/abs/astro-ph/0210603}{{\ttfamily arXiv:astro-ph/0210603
  [astro-ph]}}.

\bibitem{Seery:2005gb}
D.~Seery and J.~E. Lidsey, ``{Primordial non-Gaussianities from multiple-field
  inflation},'' \href{http://dx.doi.org/10.1088/1475-7516/2005/09/011}{{\em
  JCAP} {\bfseries 0509} (2005) 011},
\href{http://arxiv.org/abs/astro-ph/0506056}{{\ttfamily arXiv:astro-ph/0506056
  [astro-ph]}}.

\bibitem{Seery:2005wm}
D.~Seery and J.~E. Lidsey, ``{Primordial non-Gaussianities in single field
  inflation},'' \href{http://dx.doi.org/10.1088/1475-7516/2005/06/003}{{\em
  JCAP} {\bfseries 0506} (2005) 003},
\href{http://arxiv.org/abs/astro-ph/0503692}{{\ttfamily arXiv:astro-ph/0503692
  [astro-ph]}}.

\bibitem{Chen:2006nt}
X.~Chen, M.-x. Huang, S.~Kachru, and G.~Shiu, ``{Observational signatures and
  non-Gaussianities of general single field inflation},''
  \href{http://dx.doi.org/10.1088/1475-7516/2007/01/002}{{\em JCAP} {\bfseries
  0701} (2007) 002},
\href{http://arxiv.org/abs/hep-th/0605045}{{\ttfamily arXiv:hep-th/0605045
  [hep-th]}}.

\bibitem{Wands:2010af}
D.~Wands, ``{Local non-Gaussianity from inflation},''
  \href{http://dx.doi.org/10.1088/0264-9381/27/12/124002}{{\em
  Class.Quant.Grav.} {\bfseries 27} (2010) 124002},
\href{http://arxiv.org/abs/1004.0818}{{\ttfamily arXiv:1004.0818
  [astro-ph.CO]}}.

\bibitem{Chen:2010xka}
X.~Chen, ``{Primordial Non-Gaussianities from Inflation Models},''
  \href{http://dx.doi.org/10.1155/2010/638979}{{\em Adv.Astron.} {\bfseries
  2010} (2010) 638979},
\href{http://arxiv.org/abs/1002.1416}{{\ttfamily arXiv:1002.1416
  [astro-ph.CO]}}.

\bibitem{Byrnes:2014pja}
C.~T. Byrnes, ``{Lecture notes on non-Gaussianity},''
\newblock 2014.
\newblock \href{http://arxiv.org/abs/1411.7002}{{\ttfamily arXiv:1411.7002
  [astro-ph.CO]}}.
\newblock
\url{http://inspirehep.net/record/1329941/files/arXiv:1411.7002.pdf}.
\newblock

\bibitem{Ade:2013ydc}
{\bfseries Planck Collaboration} Collaboration, P.~Ade {\em et~al.}, ``{Planck
  2013 Results. XXIV. Constraints on primordial non-Gaussianity},''
\href{http://arxiv.org/abs/1303.5084}{{\ttfamily arXiv:1303.5084
  [astro-ph.CO]}}.

\bibitem{Elliston:2013afa}
J.~Elliston, D.~J. Mulryne, and R.~Tavakol, ``{What Planck does not tell us
  about inflation},'' \href{http://dx.doi.org/10.1103/PhysRevD.88.063533}{{\em
  Phys.Rev.} {\bfseries D88} (2013) 063533},
\href{http://arxiv.org/abs/1307.7095}{{\ttfamily arXiv:1307.7095
  [astro-ph.CO]}}.

\bibitem{Baumann:2014cja}
D.~Baumann, D.~Green, and R.~A. Porto, ``{B-modes and the Nature of
  Inflation},'' \href{http://dx.doi.org/10.1088/1475-7516/2015/01/016}{{\em
  JCAP} {\bfseries 1501} no.~01, (2015) 016},
\href{http://arxiv.org/abs/1407.2621}{{\ttfamily arXiv:1407.2621 [hep-th]}}.

\bibitem{Vennin:2015vfa}
V.~Vennin, K.~Koyama, and D.~Wands, ``{Encyclopaedia curvatonis},''
  \href{http://dx.doi.org/10.1088/1475-7516/2015/11/008}{{\em JCAP} {\bfseries
  1511} no.~11, (2015) 008},
\href{http://arxiv.org/abs/1507.07575}{{\ttfamily arXiv:1507.07575
  [astro-ph.CO]}}.

\bibitem{Nunes-1}
T.~S. Koivisto and N.~J. Nunes, ``{Three-form cosmology},''
  \href{http://dx.doi.org/10.1016/j.physletb.2010.01.051}{{\em Phys.Lett.}
  {\bfseries B685} (2010) 105--109},
\href{http://arxiv.org/abs/0907.3883}{{\ttfamily arXiv:0907.3883
  [astro-ph.CO]}}.

\bibitem{Nunes-2}
T.~S. Koivisto and N.~J. Nunes, ``{Inflation and dark energy from
  three-forms},'' \href{http://dx.doi.org/10.1103/PhysRevD.80.103509}{{\em
  Phys.Rev.} {\bfseries D80} (2009) 103509},
\href{http://arxiv.org/abs/0908.0920}{{\ttfamily arXiv:0908.0920
  [astro-ph.CO]}}.

\bibitem{GK09}
C.~Germani and A.~Kehagias, ``{P-nflation: generating cosmic Inflation with
  p-forms},'' \href{http://dx.doi.org/10.1088/1475-7516/2009/03/028}{{\em JCAP}
  {\bfseries 0903} (2009) 028},
\href{http://arxiv.org/abs/0902.3667}{{\ttfamily arXiv:0902.3667
  [astro-ph.CO]}}.

\bibitem{Felice-1}
A.~De~Felice, K.~Karwan, and P.~Wongjun, ``{Stability of the 3-form field
  during inflation},'' \href{http://dx.doi.org/10.1103/PhysRevD.85.123545}{{\em
  Phys.Rev.} {\bfseries D85} (2012) 123545},
\href{http://arxiv.org/abs/1202.0896}{{\ttfamily arXiv:1202.0896 [hep-ph]}}.

\bibitem{Mulryne:2012ax}
D.~J. Mulryne, J.~Noller, and N.~J. Nunes, ``{Three-form inflation and
  non-Gaussianity},''
  \href{http://dx.doi.org/10.1088/1475-7516/2012/12/016}{{\em JCAP} {\bfseries
  1212} (2012) 016},
\href{http://arxiv.org/abs/1209.2156}{{\ttfamily arXiv:1209.2156
  [astro-ph.CO]}}.

\bibitem{Koivisto:2012xm}
T.~S. Koivisto and N.~J. Nunes, ``{Coupled three-form dark energy},''
\href{http://arxiv.org/abs/1212.2541}{{\ttfamily arXiv:1212.2541
  [astro-ph.CO]}}.

\bibitem{Kumar:2014oka}
K.~S. Kumar, J.~Marto, N.~J. Nunes, and P.~V. Moniz, ``{Inflation in a two
  3-form fields scenario},''
  \href{http://dx.doi.org/10.1088/1475-7516/2014/06/064}{{\em JCAP} {\bfseries
  1406} (2014) 064},
\href{http://arxiv.org/abs/1404.0211}{{\ttfamily arXiv:1404.0211 [gr-qc]}}.

\bibitem{Barros:2015evi}
B.~J. Barros and N.~J. Nunes, ``{Three-form inflation in type II
  Randall-Sundrum},'' \href{http://dx.doi.org/10.1103/PhysRevD.93.043512}{{\em
  Phys. Rev.} {\bfseries D93} no.~4, (2016) 043512},
\href{http://arxiv.org/abs/1511.07856}{{\ttfamily arXiv:1511.07856
  [astro-ph.CO]}}.

\bibitem{Groh:2012tf}
K.~Groh, J.~Louis, and J.~Sommerfeld, ``{Duality and Couplings of
  3-Form-Multiplets in N=1 Supersymmetry},''
  \href{http://dx.doi.org/10.1007/JHEP05(2013)001}{{\em JHEP} {\bfseries 05}
  (2013) 001},
\href{http://arxiv.org/abs/1212.4639}{{\ttfamily arXiv:1212.4639 [hep-th]}}.

\bibitem{Lyth:1984gv}
D.~Lyth, ``{Large Scale Energy Density Perturbations and Inflation},''
\href{http://dx.doi.org/10.1103/PhysRevD.31.1792}{{\em Phys.Rev.} {\bfseries
  D31} (1985) 1792--1798}.

\bibitem{Sasaki1995}
M.~Sasaki and E.~D. Stewart, ``{A General analytic formula for the spectral
  index of the density perturbations produced during inflation},''
  \href{http://dx.doi.org/10.1143/PTP.95.71}{{\em Prog. Theor. Phys.}
  {\bfseries 95} (1996) 71--78},
\href{http://arxiv.org/abs/astro-ph/9507001}{{\ttfamily arXiv:astro-ph/9507001
  [astro-ph]}}.

\bibitem{Wands2000}
D.~Wands, K.~A. Malik, D.~H. Lyth, and A.~R. Liddle, ``{A New approach to the
  evolution of cosmological perturbations on large scales},''
  \href{http://dx.doi.org/10.1103/PhysRevD.62.043527}{{\em Phys. Rev.}
  {\bfseries D62} (2000) 043527},
\href{http://arxiv.org/abs/astro-ph/0003278}{{\ttfamily arXiv:astro-ph/0003278
  [astro-ph]}}.

\bibitem{Lyth2004}
D.~H. Lyth, K.~A. Malik, and M.~Sasaki, ``{A General proof of the conservation
  of the curvature perturbation},''
  \href{http://dx.doi.org/10.1088/1475-7516/2005/05/004}{{\em JCAP} {\bfseries
  0505} (2005) 004},
\href{http://arxiv.org/abs/astro-ph/0411220}{{\ttfamily arXiv:astro-ph/0411220
  [astro-ph]}}.

\bibitem{Lyth:2005du}
D.~H. Lyth and Y.~Rodriguez, ``{Non-Gaussianity from the second-order
  cosmological perturbation},''
  \href{http://dx.doi.org/10.1103/PhysRevD.71.123508}{{\em Phys.Rev.}
  {\bfseries D71} (2005) 123508},
\href{http://arxiv.org/abs/astro-ph/0502578}{{\ttfamily arXiv:astro-ph/0502578
  [astro-ph]}}.

\bibitem{Lyth:2005fi}
D.~H. Lyth and Y.~Rodriguez, ``{The Inflationary prediction for primordial
  non-Gaussianity},''
  \href{http://dx.doi.org/10.1103/PhysRevLett.95.121302}{{\em Phys.Rev.Lett.}
  {\bfseries 95} (2005) 121302},
\href{http://arxiv.org/abs/astro-ph/0504045}{{\ttfamily arXiv:astro-ph/0504045
  [astro-ph]}}.

\bibitem{Liddle-1}
A.~R. Liddle, A.~Mazumdar, and F.~E. Schunck, ``{Assisted inflation},''
  \href{http://dx.doi.org/10.1103/PhysRevD.58.061301}{{\em Phys.Rev.}
  {\bfseries D58} (1998) 061301},
\href{http://arxiv.org/abs/astro-ph/9804177}{{\ttfamily arXiv:astro-ph/9804177
  [astro-ph]}}.

\bibitem{Starobinsky:1986fxa}
A.~A. Starobinsky, ``{Multicomponent de Sitter (Inflationary) Stages and the
  Generation of Perturbations},'' {\em JETP Lett.} {\bfseries 42} (1985)
  152--155.
[Pisma Zh. Eksp. Teor. Fiz.42,124(1985)].

\bibitem{Kenton:2015lxa}
Z.~Kenton and D.~J. Mulryne, ``{The squeezed limit of the bispectrum in
  multi-field inflation},''
  \href{http://dx.doi.org/10.1088/1475-7516/2015/10/018}{{\em JCAP} {\bfseries
  1510} no.~10, (2015) 018},
\href{http://arxiv.org/abs/1507.08629}{{\ttfamily arXiv:1507.08629
  [astro-ph.CO]}}.

\bibitem{Kenton:2016abp}
Z.~Kenton and D.~J. Mulryne, ``{The Separate Universe Approach to Soft
  Limits},''
\href{http://arxiv.org/abs/1605.03435}{{\ttfamily arXiv:1605.03435
  [astro-ph.CO]}}.

\bibitem{Gao:2008dt}
X.~Gao, ``{Primordial Non-Gaussianities of General Multiple Field Inflation},''
  \href{http://dx.doi.org/10.1088/1475-7516/2008/06/029}{{\em JCAP} {\bfseries
  0806} (2008) 029},
\href{http://arxiv.org/abs/0804.1055}{{\ttfamily arXiv:0804.1055 [astro-ph]}}.

\bibitem{Vernizzi:2006ve}
F.~Vernizzi and D.~Wands, ``{Non-gaussianities in two-field inflation},''
  \href{http://dx.doi.org/10.1088/1475-7516/2006/05/019}{{\em JCAP} {\bfseries
  0605} (2006) 019},
\href{http://arxiv.org/abs/astro-ph/0603799}{{\ttfamily arXiv:astro-ph/0603799
  [astro-ph]}}.

\bibitem{felice2}
A.~De~Felice, K.~Karwan, and P.~Wongjun, ``{Reheating in 3-form inflation},''
  \href{http://dx.doi.org/10.1103/PhysRevD.86.103526}{{\em Phys.Rev.}
  {\bfseries D86} (2012) 103526},
\href{http://arxiv.org/abs/1209.5156}{{\ttfamily arXiv:1209.5156
  [astro-ph.CO]}}.

\bibitem{Rigopoulos:2003ak}
G.~I. Rigopoulos and E.~P.~S. Shellard, ``{The separate universe approach and
  the evolution of nonlinear superhorizon cosmological perturbations},''
  \href{http://dx.doi.org/10.1103/PhysRevD.68.123518}{{\em Phys. Rev.}
  {\bfseries D68} (2003) 123518},
\href{http://arxiv.org/abs/astro-ph/0306620}{{\ttfamily arXiv:astro-ph/0306620
  [astro-ph]}}.

\bibitem{Christopherson:2008ry}
A.~J. Christopherson and K.~A. Malik, ``{The non-adiabatic pressure in general
  scalar field systems},''
  \href{http://dx.doi.org/10.1016/j.physletb.2009.04.003}{{\em Phys. Lett.}
  {\bfseries B675} (2009) 159--163},
\href{http://arxiv.org/abs/0809.3518}{{\ttfamily arXiv:0809.3518 [astro-ph]}}.

\bibitem{Byrnes:2008wi}
C.~T. Byrnes, K.-Y. Choi, and L.~M.~H. Hall, ``{Conditions for large
  non-Gaussianity in two-field slow-roll inflation},''
  \href{http://dx.doi.org/10.1088/1475-7516/2008/10/008}{{\em JCAP} {\bfseries
  0810} (2008) 008},
\href{http://arxiv.org/abs/0807.1101}{{\ttfamily arXiv:0807.1101 [astro-ph]}}.

\bibitem{Peterson:2010np}
C.~M. Peterson and M.~Tegmark, ``{Testing Two-Field Inflation},''
  \href{http://dx.doi.org/10.1103/PhysRevD.83.023522}{{\em Phys.Rev.}
  {\bfseries D83} (2011) 023522},
\href{http://arxiv.org/abs/1005.4056}{{\ttfamily arXiv:1005.4056
  [astro-ph.CO]}}.

\bibitem{Elliston:2011dr}
J.~Elliston, D.~J. Mulryne, D.~Seery, and R.~Tavakol, ``{Evolution of fNL to
  the adiabatic limit},''
  \href{http://dx.doi.org/10.1088/1475-7516/2011/11/005}{{\em JCAP} {\bfseries
  1111} (2011) 005},
\href{http://arxiv.org/abs/1106.2153}{{\ttfamily arXiv:1106.2153
  [astro-ph.CO]}}.

\bibitem{Elliston:2012wm}
J.~Elliston, L.~Alabidi, I.~Huston, D.~Mulryne, and R.~Tavakol, ``{Large
  trispectrum in two-field slow-roll inflation},''
  \href{http://dx.doi.org/10.1088/1475-7516/2012/09/001}{{\em JCAP} {\bfseries
  1209} (2012) 001},
\href{http://arxiv.org/abs/1203.6844}{{\ttfamily arXiv:1203.6844
  [astro-ph.CO]}}.

\bibitem{Mollerach:1989hu}
S.~Mollerach, ``{Isocurvature baryon perturbations and inflation},''
\href{http://dx.doi.org/10.1103/PhysRevD.42.313}{{\em Phys.Rev.} {\bfseries
  D42} (1990) 313--325}.

\bibitem{Linde:1996gt}
A.~D. Linde and V.~F. Mukhanov, ``{Nongaussian isocurvature perturbations from
  inflation},'' \href{http://dx.doi.org/10.1103/PhysRevD.56.R535}{{\em
  Phys.Rev.} {\bfseries D56} (1997) 535--539},
\href{http://arxiv.org/abs/astro-ph/9610219}{{\ttfamily arXiv:astro-ph/9610219
  [astro-ph]}}.

\bibitem{Enqvist:2001zp}
K.~Enqvist and M.~S. Sloth, ``{Adiabatic CMB perturbations in pre - big bang
  string cosmology},''
  \href{http://dx.doi.org/10.1016/S0550-3213(02)00043-3}{{\em Nucl.Phys.}
  {\bfseries B626} (2002) 395--409},
\href{http://arxiv.org/abs/hep-ph/0109214}{{\ttfamily arXiv:hep-ph/0109214
  [hep-ph]}}.

\bibitem{Lyth:2001nq}
D.~H. Lyth and D.~Wands, ``{Generating the curvature perturbation without an
  inflaton},'' \href{http://dx.doi.org/10.1016/S0370-2693(01)01366-1}{{\em
  Phys.Lett.} {\bfseries B524} (2002) 5--14},
\href{http://arxiv.org/abs/hep-ph/0110002}{{\ttfamily arXiv:hep-ph/0110002
  [hep-ph]}}.

\bibitem{Moroi:2001ct}
T.~Moroi and T.~Takahashi, ``{Effects of cosmological moduli fields on cosmic
  microwave background},''
  \href{http://dx.doi.org/10.1016/S0370-2693(01)01295-3}{{\em Phys.Lett.}
  {\bfseries B522} (2001) 215--221},
\href{http://arxiv.org/abs/hep-ph/0110096}{{\ttfamily arXiv:hep-ph/0110096
  [hep-ph]}}.

\bibitem{Enqvist:2005pg}
K.~Enqvist and S.~Nurmi, ``{Non-gaussianity in curvaton models with nearly
  quadratic potential},''
  \href{http://dx.doi.org/10.1088/1475-7516/2005/10/013}{{\em JCAP} {\bfseries
  0510} (2005) 013},
\href{http://arxiv.org/abs/astro-ph/0508573}{{\ttfamily arXiv:astro-ph/0508573
  [astro-ph]}}.

\bibitem{Linde:2005yw}
A.~D. Linde and V.~Mukhanov, ``{The curvaton web},''
  \href{http://dx.doi.org/10.1088/1475-7516/2006/04/009}{{\em JCAP} {\bfseries
  0604} (2006) 009},
\href{http://arxiv.org/abs/astro-ph/0511736}{{\ttfamily arXiv:astro-ph/0511736
  [astro-ph]}}.

\bibitem{Malik:2006pm}
K.~A. Malik and D.~H. Lyth, ``{A numerical study of non-gaussianity in the
  curvaton scenario},''
  \href{http://dx.doi.org/10.1088/1475-7516/2006/09/008}{{\em JCAP} {\bfseries
  0609} (2006) 008},
\href{http://arxiv.org/abs/astro-ph/0604387}{{\ttfamily arXiv:astro-ph/0604387
  [astro-ph]}}.

\bibitem{Sasaki:2006kq}
M.~Sasaki, J.~Valiviita, and D.~Wands, ``{Non-Gaussianity of the primordial
  perturbation in the curvaton model},''
  \href{http://dx.doi.org/10.1103/PhysRevD.74.103003}{{\em Phys.Rev.}
  {\bfseries D74} (2006) 103003},
\href{http://arxiv.org/abs/astro-ph/0607627}{{\ttfamily arXiv:astro-ph/0607627
  [astro-ph]}}.

\bibitem{Meyers:2013gua}
J.~Meyers and E.~R.~M. Tarrant, ``{Perturbative Reheating After Multiple-Field
  Inflation: The Impact on Primordial Observables},''
  \href{http://dx.doi.org/10.1103/PhysRevD.89.063535}{{\em Phys.Rev.}
  {\bfseries D89} (2014) 063535},
\href{http://arxiv.org/abs/1311.3972}{{\ttfamily arXiv:1311.3972
  [astro-ph.CO]}}.

\bibitem{Elliston:2014zea}
J.~Elliston, S.~Orani, and D.~J. Mulryne, ``{General analytic predictions of
  two-field inflation and perturbative reheating},''
  \href{http://dx.doi.org/10.1103/PhysRevD.89.103532}{{\em Phys.Rev.}
  {\bfseries D89} (2014) 103532},
\href{http://arxiv.org/abs/1402.4800}{{\ttfamily arXiv:1402.4800
  [astro-ph.CO]}}.

\bibitem{Byrnes:2014xua}
C.~T. Byrnes, M.~Cortês, and A.~R. Liddle, ``{Comprehensive analysis of the
  simplest curvaton model},''
  \href{http://dx.doi.org/10.1103/PhysRevD.90.023523}{{\em Phys. Rev.}
  {\bfseries D90} no.~2, (2014) 023523},
\href{http://arxiv.org/abs/1403.4591}{{\ttfamily arXiv:1403.4591
  [astro-ph.CO]}}.

\bibitem{Fergusson:2008ra}
J.~R. Fergusson and E.~P.~S. Shellard, ``{The shape of primordial
  non-Gaussianity and the CMB bispectrum},''
  \href{http://dx.doi.org/10.1103/PhysRevD.80.043510}{{\em Phys. Rev.}
  {\bfseries D80} (2009) 043510},
\href{http://arxiv.org/abs/0812.3413}{{\ttfamily arXiv:0812.3413 [astro-ph]}}.

\bibitem{Fergusson:2009nv}
J.~R. Fergusson, M.~Liguori, and E.~P.~S. Shellard, ``{General CMB and
  Primordial Bispectrum Estimation I: Mode Expansion, Map-Making and Measures
  of $f_{NL}$},'' \href{http://dx.doi.org/10.1103/PhysRevD.82.023502}{{\em
  Phys. Rev.} {\bfseries D82} (2010) 023502},
\href{http://arxiv.org/abs/0912.5516}{{\ttfamily arXiv:0912.5516
  [astro-ph.CO]}}.

\bibitem{Babich:2004gb}
D.~Babich, P.~Creminelli, and M.~Zaldarriaga, ``{The Shape of
  non-Gaussianities},''
  \href{http://dx.doi.org/10.1088/1475-7516/2004/08/009}{{\em JCAP} {\bfseries
  0408} (2004) 009},
\href{http://arxiv.org/abs/astro-ph/0405356}{{\ttfamily arXiv:astro-ph/0405356
  [astro-ph]}}.

\end{thebibliography}\endgroup

\end{document}